\documentclass[a4paper,fleqn,authoryear]{cas-dc}

\ExplSyntaxOn
\cs_gset:Npn \__first_footerline:
  {
    \group_begin:
      \small \sffamily
      \ifnum \theblind > 0\relax
      \else
        \__short_authors: 
      \fi
      {\rmfamily \itshape Preprint} 
    \group_end:
  }
\ExplSyntaxOff

\makeatletter
\@ifundefined{orcid}{}{\renewcommand{\orcid}[1]{}}        
\@ifundefined{printorcid}{}{} 
\@ifundefined{orcidicon}{}{}   
\makeatother

\usepackage[authoryear]{natbib}
\usepackage{cuted} 
\usepackage{titlesec}
\usepackage{bbm}
\usepackage{tabularx}
\usepackage{graphicx}
\usepackage{booktabs}
\usepackage{siunitx}
\newcolumntype{Y}{S[table-format=2.2(2)]} 

\newcommand{\best}[1]{\textbf{#1}}

\def\tsc#1{\csdef{#1}{\textsc{\lowercase{#1}}\xspace}}
\tsc{WGM}
\tsc{QE}
\tsc{EP}
\tsc{PMS}
\tsc{BEC}
\tsc{DE}

\titleformat{\section}
  {\large\bfseries}{\thesection.}{0.6em}{}

\titleformat{\subsection}
  {\normalsize\itshape}{\thesubsection.}{0.6em}{}

\titleformat{\subsubsection}
  {\normalsize\itshape}{\thesubsubsection.}{0.6em}{}

\titleformat{\section}
  {\large\bfseries}{\thesection.}{0.6em}{}
\titleformat{\subsection}
  {\normalsize\itshape}{\thesubsection.}{0.6em}{}
\titlespacing*{\section}{0pt}{*2}{*0.6}
\titlespacing*{\subsection}{0pt}{*1.4}{*1.4}

\usepackage{graphicx}    
\usepackage{subcaption}  
\usepackage{enumitem}

\usepackage{amsmath,amssymb}   

\usepackage{algorithm}         
\usepackage{algxpar}  

\begin{document}
\let\WriteBookmarks\relax
\def\floatpagepagefraction{1}
\def\textpagefraction{.001}

\shorttitle{}

\shortauthors{}

\title [mode = title]{A Semi Centralized Training Decentralized Execution Architecture for Multi-Agent Deep Reinforcement Learning in Traffic Signal Control}


%

\author[1]{Arash Rezaali}
\fnmark[1]

\author[1]{Pouria Yazdani}
\fnmark[1]

\author[1]{Monireh Abdoos}
\cormark[1]

\fntext[1]{Co-first authors. These authors contributed equally to this work and are listed in alphabetical order.}

\cortext[1]{Corresponding author.\\
\textit{E-mail address:} m\_abdoos@sbu.ac.ir (M.~Abdoos). 
\newline}

\affiliation[1]{organization={Faculty of Computer Science and Engineering, Shahid Beheshti University},
    city={Tehran},
    country={Iran}}
    


\begin{abstract}
Multi-agent reinforcement learning (MARL) has emerged as a promising paradigm for adaptive traffic signal control (ATSC) of multiple intersections. Existing approaches typically follow either a fully centralized or a fully decentralized design. Fully centralized approaches suffer from the curse of dimensionality, and reliance on a single learning server, whereas purely decentralized approaches operate under severe partial observability and lack explicit coordination resulting in suboptimal performance. These limitations motivate region-based MARL, where the network is partitioned into smaller, tightly Interdependent intersections that form regions, and training is organized around these regions. This paper introduces a \emph{Semi-Centralized Training, Decentralized Execution} (SEMI-CTDE) architecture for multi intersection ATSC. Within each region, SEMI-CTDE performs centralized training with regional parameter sharing and employs composite state and reward formulations that jointly encode local and regional information. The architecture is highly transferable across different policy backbones and state–reward instantiations. Building on this architecture, we implement two models with distinct design objectives. A multi-perspective experimental analysis of the two implemented SEMI-CTDE-based models covering ablations of the architecture's core elements including rule based and fully decentralized baselines shows that they achieve consistently superior performance and remain effective across a wide range of traffic densities and distributions. 
\end{abstract}




\maketitle

\section{Introduction}

\hspace*{1em}Traffic congestion is a major and complex challenge for cities worldwide with the rapid growth of urbanization and vehicle ownership. Studies show that heavy traffic conditions raise fuel consumption and increase commute times \citep{YANG2023101431}. For instance, according to the 2024 INRIX Global Traffic Scorecard, individual commuters in Istanbul, New York City, and Chicago experienced total annual delay of about 105, 102, and 102 hours, respectively, underscoring the magnitude of intersection-driven delays in major metros  \citep{inrix2024scorecard}. In urban road networks, currently implemented signal control policies at signalized intersections are often inefficient, making them a major source of congestion, increased delays, queues, and elevated vehicle emissions \citep{Eom2020TrafficSignalControl}.

Reinforcement learning (RL) has become a standard solution for adaptive traffic signal control (ATSC), controlling phase selection and timing as a sequential decision problem that optimizes long-horizon objectives such as delay, throughput, and emissions under nonstationary demand \citep{yau2017survey}.  Deep RL (DRL) extends this approach by using function approximation to digest rich state representations, ranging from detector queues to trajectories and graph-structured networks. This capability enables the implementation of policies that generalize across varying traffic flows and network topologies \citep{Zhao2024}. Collectively, this body of work motivates moving beyond single-intersection controllers toward coordinated, network-level solutions and setting the stage for multi-agent formulations.

Multi-agent deep reinforcement learning (MADRL) \citep{HU2024128068} models each intersection as an intelligent agent. This allows the policies to adapt to local, nonstationary demand. Additionally, it learns coordination mechanisms across the network. MADRL is particularly well‑suited for ATSC because intersections act as distributed decision makers. Their actions jointly shape traffic flow and dynamics throughout the network, which propagate benefits downstream. By enabling jointly coordinated actions among neighboring agents, MADRL captures the interdependent structure of real networks more accurately than Independent RL (IRL) formulations \citep{liang2025survey}.

However, applying MADRL to traffic networks presents several key challenges. Traffic flow dynamics are inherently interdependent, as the chosen action of an agent at one intersection influences the traffic congestion at neighboring intersections. This results in cascade effect, where the queue discharge from one intersection becomes the inflow for downstream intersections. This interaction creates complex spatial-temporal dependencies across the entire network. If each agent optimizes signal timings independently, it can lead to suboptimal or even adverse emergent behaviors (e.g., one intersection starving its neighbors or causing spillback). The multi-agent setting is partially observable from any single intersection’s perspective. Each agent has a limited local view, and the environment’s dynamics become non-stationary as all agents learn concurrently. Ideally, a fully centralized controller would coordinate all intersections jointly, explicitly accounting for the state and flows of every other agent to optimize global performance \citep{Casas2017}. Yet in practice it is computationally intractable and not scalable at city scale due to the huge joint state and action spaces. It also imposes significant latency and communication overhead \citep{8667868}. Conversely, a purely decentralized approach where each traffic light learns on its own lacks global optimality and may fail to resolve traffic network-level congestion patterns \citep{Bao2023}. In summary, pure centralized and pure decentralized solutions each have drawbacks. This observation has driven researchers towards hybrid approaches that balance coordination and scalability, namely distributed-training decentralized-execution frameworks, covering a spectrum of training paradigms that vary in their degree of centralization, from fully centralized to fully decentralized \citep{Noaeen2022,saadi2025survey}.

Among these, region-based multi-agent approaches have gained attention as a practical compromise between fully centralized and fully decentralized control. Partitioning a large network into smaller regions of closely interacting intersections, can reduce complexity and improve local coordination. Recent studies have shown that dividing a traffic network into strongly interdependent regions and solving regional sub-problems can significantly speed up learning and improve performance compared to tackling the entire traffic network as one problem \citep{10637352}. Within these set of approaches, the joint design of state and reward is essential for region-based multi-agent approaches because each intersection must reason along two complementary perspectives: a local context (e.g., approach-level queues) and a region context that abstracts interdependent dynamics across same-region neighbors.

 Our research thoroughly investigates how a region-based MADRL architecture can operate effectively under this joint design while avoiding the drawbacks of fully centralized and fully decentralized training paradigms. We argue that partitioning the traffic network into tightly interdependent regions of intersections and then applying centralized learning within each region using composite states that capture both local and regional information, and composite rewards that jointly optimize local throughput and regional coordination substantially enhances overall traffic flow efficiency in urban networks. To make these ideas practical, we propose a structured architecture that generalizes across alternative instantiations, allowing regional based methods to be implemented with different policy backbones and state and reward definitions. We refer to this architecture as SEMI-CTDE : \textbf{Semi}-\textbf{C}entralized \textbf{T}raining, \textbf{D}ecentralized \textbf{E}xecution, which we primarily study in the context of urban traffic signal control.

This paper’s main contributions are summarized as follows:

\begin{itemize}

    \item Modular SEMI-CTDE architecture using MADRL. We propose a region-based SEMI-CTDE architecture for TSC that jointly designs composite state and reward definitions together with their dedicated feature blocks, ensuring that both local and regional aspects of traffic dynamics are captured. The architecture is highly transferable, allowing different policy backbones, feature extraction methods, and region formation algorithms to be substituted without changing the underlying conceptual design.
  \item Two realized implementations.
  We instantiate the proposed architecture in two models with distinct design in regional objectives, providing concrete examples of how the SEMI-CTDE architecture can be applied in practice.

  \item Comprehensive and multi-perspective experiments.
Core principles of the proposed architecture are assessed through targeted ablation studies and evaluations across diverse traffic conditions, yielding clear relationships between region partitioning, composite state/reward design and different design objectives under varying demand conditions.

\end{itemize}

\section{Related Works}
\hspace*{1em} RL and MARL have been applied in addressing ATSC problems in recent years, due to their strong empirical performance, and ability to adapt to diverse traffic conditions \citep{yau2017survey}. The review in this section begins with RL methods designed for single-intersection environments. Then it transitions to multi-intersection networks, covering both centralized and decentralized MARL for ATSC. We further review methods for extracting regions from urban traffic networks and survey region-based MARL frameworks proposed for adaptive traffic signal control.

\subsection{Reinforcement Learning for Traffic Signal control}
\hspace*{1em} Early work on ATSC model a single intersection as a Markov Decision Process and applied tabular RL, most notably Q-learning \citep{Watkins1992}, to learn mappings from compact state descriptors (e.g., current phase and approach queues) to discrete phase-change actions through trial and error \citep{Abdulhai2003}. While these methods demonstrated that RL can outperform fixed-time and actuated control in simulation, they relied on coarse state discretization, small action spaces, and carefully engineered features. As a result, these methods cannot adequately represent rich state spaces, generalize across different traffic conditions, and scale effectively as the number of phases and lanes increases \citep{Noaeen2022}.

DRL approaches for single-intersection control address many of these limitations by learning value or policy approximators over higher-dimensional representations \citep{Rasheed2020}. The method proposed in \citet{Liang2019} formulates the intersection as a grid-based state constructed from sensor data and use a dueling double-DQN with prioritized replay to adapt cycle lengths and phase durations, achieving lower delay and shorter queues than classical plans. In \citet{Ault2020}, an interpretable DRL-based precedence-function controller and three DQN variants that tune a polynomial control function is developed, showing that such a regulatable policy can match DNN-based controllers. These works illustrate how DRL can exploit richer state spaces with consistent reward function formulations and more expressive policies than tabular RL. This domain has been widely studied and surveyed in recent years \citep{Zhao2024,saadi2025survey,Bouktif2023}. While single-intersection RL controllers can improve local performance, real-world urban networks invariably include multiple interacting intersections whose dynamics are mutually dependent, making multi-agent RL formulations essential for capturing these dependencies and coordinating control decisions across the network.

\subsection{Multi-Agent Deep Reinforcement Learning}
\hspace*{1em} For multi-intersection adaptive control, the most direct extension of single-agent DRL is to treat the whole network as a single global agent, yielding a centralized-training centralized-execution (CTCE) paradigm. In this setting a single deep network receives a global state that aggregates measurements from all intersections and outputs a joint action vector specifying the signal timings network-wide. A representative example is the spatial–temporal DRL model of  \citet{9922459}, which encodes the entire network’s traffic conditions into a spatio-temporal feature representation and computes signal plans for all junctions within one centralized controller. Such CTCE paradigm can in principle exploit global information to coordinate signals, but they quickly encounter the curse of dimensionality in both the state and action spaces as the network grows.

To improve scalability while retaining some form of centralized guidance, many recent works adopt a centralized-training–decentralized-execution (CTDE) paradigm, where each intersection is an agent with its own policy, but training is assisted by a centralized critic or server. One area of study employs global critics or value-decomposition structures: for example, in \citet{BIE2024104663}, a collaborative MARL method is proposed with a spatio-temporal graph attention network and value decomposition to learn a joint value function over heterogeneous intersections with decentralized policies at execution time; and \citet{SONG2024104528} use a counterfactual multi-agent actor–critic approach in which a centralized critic leverages joint observations to compute counterfactual advantages for each intersection actor. Another area introduces explicit parameter-server or federated mechanisms: \citet{REN2024121111} propose a two-layer coordinated RL approach where local agents send experience to a central learner that optimizes a coordination policy and broadcasts updated parameters back to the intersections, while \citet{Li2025} design a federated DRL framework in which intersection agents periodically upload locally trained models to a central server for aggregation and then receive a global model for continued training and decentralized execution. Other CTDE approaches emphasize neighborhood structure in the agent design: \citet{CAI2025126938} models each intersection as an interpretable multi-agent RL controller that conditions on local and neighboring information, yet still relies on a central learner that aggregates experience across agents during training while policies are executed locally at each intersection.

A third paradigm pursues fully decentralized training and execution (DTDE), where there is no central critic or server and both learning and control are carried out in a distributed manner. Early work by \citet{6083114} applies independent multi-agent Q-learning, training a separate tabular Q-learner at each intersection using only local state and reward, without any explicit coordination mechanism. More recent methods introduce local cooperation via peer-to-peer communication or consensus: \citet{LIU2022390} propose a distributed DRL method in which each intersection trains a local deep controller and uses a consensus algorithm over the communication graph to align policies without a central learner; \citet{LIU2025130834}’s decentralized neighboring information fusion (D-NIF) algorithm further develops this idea by letting agents fuse neighbors’ information and exchange parameters through an extra-gradient consensus step, achieving both decentralized modeling and decentralized distributed training; and \citet{9103316} design a multi-agent DRL method for urban traffic light control in networks where controllers are deployed at intersections and coordinate via local information exchange within the communication network, again without a global training server. Together, these CTCE, CTDE, and DTDE intersection-level MARL paradigms illustrate the spectrum from fully centralized to fully decentralized designs, and motivate region-level MARL architectures that seek a middle ground between global coordination and scalability.

\subsection{Region-based MARL for TSC}
\hspace*{1em} Region-based MARL approaches extend these ideas by introducing an intermediate spatial scale: instead of controlling each intersection independently or aggregating the whole network into a single agent, the network is partitioned into regions of tightly interdependent intersections and RL controllers are defined at the region level. The effectiveness of such approaches depends critically on how these regions are constructed. Beyond the RL literature, several works in traffic networks have studied how to partition heterogeneous networks into internally homogeneous, strongly interacting regions; for example \citet{SAEEDMANESH2016250} propose a clustering procedure that groups links into macroscopic-fundamental-diagram-consistent regions based on directional flow patterns, while \citet{su14169802} design a dynamic regional partitioning method for active traffic control that updates control subareas according to evolving congestion and correlation patterns. These ideas motivate the use of data-driven zoning as a basis for region-level reinforcement learning.

One set of approaches keep a largely centralized perspective and uses regions primarily as a decomposition of the global problem. The CODER framework of \citet{8676356} partitions a large traffic grid into several subregions with identical topology, trains a DRL agent for each subregion, and then introduces a centralized global agent that aggregates regional value estimates into a single global $Q$-function from which the joint action over all regions is selected. Training and execution therefore still pass through a global coordinator, and the region abstraction mainly serves to reduce the dimensionality and facilitate reuse of subregion policies rather than to fully decentralize control.

In contrast, region-level CTDE methods assign a distinct learning agent to each region and rely on centralized or shared components only during training. RegionSTLight \citep{10637352} derives a regional multi-agent Q-learning framework in which the global $Q$-value is decomposed into a sum of regional $Q$-values and combines this with a dynamic zoning algorithm that groups intersections into strongly interdependent regions on the basis of real-time link flow densities; a lightweight spatio–temporal fusion network encodes intra-region interactions, while the value-decomposition structure provides centralized guidance during training and execution remains distributed across regions. Building on these methods, \citet{10966978} augment regional MARL architectures such as RegionLight and Regional-DRL with GA2 communication modules that aggregate macro- and micro-level traffic states across regions; although control actions are executed by regional agents, parameter updates depend on globally aggregated information, preserving the CTDE character at the region scale.

Several region-level DTDE approaches push decentralization further by letting each region learn independently from its own experience. The fuzzy-graph method of \citet{9109698} first extracts correlated sets of intersections as regions using an $\alpha$-cut on a fuzzy relation graph and then learns a single Q-learning controller per region on regional states and rewards, without any global critic. RegionLight \citep{10490249} formulates a constrained network-partitioning problem that produces star-topology regions and trains an adaptive deep RL controller in each region independently, while \citep{LU2025} employ a regional soft actor–critic approach in a connected-vehicle environment where SAC agents control predefined regions using local and neighboring information. Together, these works show that region-based MARL can be realized along the same CTCE–CTDE–DTDE spectrum as intersection-level methods, while region formation and region-level state and reward design play a central role in balancing scalability, decentralization, and coordination.

Overall, the literature shows that, despite substantial progress, there is still no unified, region-centric MADRL architecture that effectively balances coordination, scalability, and practical deployability for urban TSC. These gaps motivate us to develop and empirically test a SEMI-CTDE architecture that is explicitly region-centric, couples principled region formation with jointly designed composite state and reward representations, and centralizes learning only within tightly interdependent regions while preserving decentralized execution at the intersection level.
\section{Problem Definition}

\hspace*{1em}In this section, we outline RL-based modeling of TSC and then establish the problem formulation for region-based MARL in TSC. Key notations used throughout this paper are summarized in Table~\ref{tab:notation}.
\begin{table}[t]
\centering
\caption{Summary of key notations.}
\label{tab:notation}
\resizebox{0.98\columnwidth}{!}{%
\begin{tabular}{ll}
\toprule
\textbf{Notation} & \textbf{Description} \\
$I$                   & Set of all intersection agents in the network \\
$R_k$                           & Region $k$  \\
$\mathcal{R}(i)$                          & Region-assignment mapping for intersection agent $i$ \\
$K$                             & Number of regions in the partition \\
$U$                             & Set of regional agent sets \\
$U_k$                           & Set of regional agents associated with region $k$ \\
$u_i$                          & Regional agent assigned to intersection agent $i$ \\
$\mathcal{A}^G$                 & Global action space  \\
$\mathcal{A}_i$                 & Admissible action set of intersection agent $i$ \\
$a_i^{(n)}$                     & Action of intersection agent $i$ at decision step $n$ \\
$t_i^{(n)}$                     & Time of the $n$-th decision at intersection agent $i$ \\
$S_i$                           & State space of intersection agent $i$ \\
$r_i(t_i^{(n)})$                     & Reward assigned to intersection agent $i$ at decision step $n$ \\
$s_i^{\text{local}}$            & Local component of composite state at intersection agent $i$ \\
$s_i^{\text{regional}}$         & Regional component of composite state at intersection agent $i$ \\

$\boldsymbol{\phi}_i^{p}$      & Local phase context feature at intersection agent $i$ \\
$\boldsymbol{\phi}_i^{t}$ & Local throughput feature at intersection agent $i$ \\
$\boldsymbol{\phi}_i^{s}$    & Local spatial feature at intersection agent $i$ \\

$\boldsymbol{\psi}_i^{p}$        & Regional phase context feature at intersection agent $i$ \\
$\boldsymbol{\psi}_i^{t}$   & Regional throughput feature at intersection agent $i$ \\
$\boldsymbol{\psi}_i^{s}$      & Regional spatial feature at intersection agent $i$ \\
$r_i^{\text{local}}(t_i^{(n)})$                     & Local reward term of intersection agent $i$ at decision step $n$ \\
$r_i^{\text{regional}}(t_i^{(n)})$                  & Regional reward term of intersection agent $i$ at decision step $n$ \\
$\beta_{\mathcal{R}(i)}$                & Weight on $r_i^{\text{local}}$ in region $\mathcal{R}(i)$ \\
$1-\beta_{\mathcal{R}(i)}$             & Weight on $r_i^{\text{regional}}$ in region $\mathcal{R}(i)$ \\
$\mathcal{D}_{R_k}$           & Region-level shared replay memory for region $R_k$ \\
$\theta_k$              & Parameters of the regional DDQN for region $R_k$ \\
$\bar{\theta}_k$        & Target-network parameters of the regional DDQN for region $R_k$ \\
$\mathcal{E}_{R(i)}$                   & Set of all approaches belonging to region $\mathcal{R}(i)$ \\
$\tau$                                 & Spillback threshold for halted vehicles on an approach \\
$\mathcal{E}^{\text{boundary}}_{R(i)}$ & Set of boundary approaches of $\mathcal{R}(i)$ \\
$N_i^{\text{hop}}$                    & Ordered set of directional OneHop neighbors of intersection agent $i$ \\

$g_s$ & Short green phase duration \\
$g_l$ & Long green phase duration \\

\bottomrule
\end{tabular}%
}
\end{table}

\subsection{RL-based Modeling of Traffic Signal Control Problem}
\label{sec:rl-based-modeling-of-tsc-problem}

\hspace*{1em}A standard way to model the problem of ATSC is to model each signalized intersection as a Markov Decision Process (MDP) that repeatedly maps observed traffic conditions into signal control actions. Formally, at each decision time $t$, an intersection $i$ is represented as $(s_{t}, a_{t}, P, r_{t}, \gamma)$: the \emph{state} $s_t \in S$ captures local traffic conditions that are sufficient to approximate the Markov property \citep{SuttonBarto2018}; the \emph{action} $a_t \in \mathcal{A}$ selects the next admissible signal operation; the \emph{transition function} $P(s_{t+1} \mid s_t, a_t)$ is determined by stochastic arrivals and network flow dynamics; and the \emph{reward} $r_t = r(s_t, a_t)$ reflects objectives such as reducing delay and queues.

The control goal is to choose a \emph{policy} $\pi(a \mid s)$ that maximizes the expected discounted return:
\begin{equation}
\label{eq:control-goal-rl}
    J(\pi) = \mathbb{E}_{\pi} \left[ \sum_{t=0}^{\infty} \gamma^{t} \, r(s_t, a_t) \right], \quad \gamma \in (0,1),
\end{equation}
where the expectation is taken over trajectories generated by $s_{t+1} \sim P(\cdot \mid s_t, a_t)$ and $a_t \sim \pi(\cdot \mid s_t)$. Here, the discount factor $\gamma$ down-weights future rewards, induces an effective horizon $H_{\mathrm{eff}} \approx (1 - \gamma)^{-1}$, and (for $0 < \gamma < 1$) makes the Bellman operator a contraction. Choosing $\gamma$ close to $1$ emphasizes long-term congestion mitigation, whereas smaller $\gamma$ prioritizes immediate delay reductions.

This formulation captures the core tension in TSC. Phase decisions at each step affect not only instantaneous delay but also future congestion through $P$ and $\gamma$. Modeling each intersection as an MDP thus provides a principled foundation to reason about sequential trade-offs under uncertainty.

\emph{Q-learning} \citep{Watkins1992} is a method that implements the objective in Eq. \eqref{eq:control-goal-rl} by learning the optimal action-value function $Q^{*}(s,a)$ directly. Given Eq. \eqref{eq:control-goal-rl}, value functions satisfy the Bellman relations:
\begin{equation}
\begin{aligned}
V^{\pi}(s) &= \mathbb{E}\!\left[r(s,a) + \gamma \, V^{\pi}(s') \mid s\right] \\
Q^{\pi}(s,a) &= \mathbb{E}\!\left[r(s,a) + \gamma \, \mathbb{E}_{a'\sim\pi(\cdot\mid s')} Q^{\pi}(s',a')\right]
\end{aligned}
\end{equation}

\noindent
Here, $V^{\pi}(s)$ is the \emph{state-value} under policy $\pi$, and $Q^{\pi}(s,a)$ is the \emph{action-value}; $r(s,a)$ is the instantaneous or delayed reward obtained when the \emph{agent} takes action $a$ after observing state $s$; $s'$ denotes the \emph{next observation/state}; and the nested expectation $\mathbb{E}_{a'\sim\pi(\cdot\mid s')}$ expresses that the next action is \emph{chosen} according to policy $\pi$. The outer expectation is with respect to the stochastic transition $s'\sim P(\cdot\mid s,a)$. The optimal counterpart obeys the Bellman optimality equation:
\begin{equation}
Q^{*}(s,a) = \mathbb{E}\!\left[r(s,a) + \gamma \, \max_{a'\in \mathcal{A}} Q^{*}(s',a')\right].
\end{equation}
Q-learning updates a tabular estimate toward this fixed point using sampled transitions $(s_t,a_t,r_t,s_{t+1})$:
{\setlength{\mathindent}{0pt}
\begin{equation}
\begin{aligned}
&\text{\ \ \ \ \ \ \ \ \ \ }Q_{t+1}(s_t,a_t) \leftarrow Q_t(s_t,a_t) + \eta\,\delta_t\\
&\text{\ \ \ \ \ \ \ \ \ \ }\delta_t = r_t + \gamma \max_{a'\in \mathcal{A}} Q_t(s_{t+1},a') - Q_t(s_t,a_t)
\end{aligned}
\end{equation}
}

\noindent
where $\delta_t$ is the \emph{temporal-difference (TD) error} between the target and current estimate; $\eta\in(0,1]$ is the \emph{learning rate}; and $(s_t,a_t,r_t,s_{t+1})$ are sampled by the agent while interacting with the environment; action selection balances exploration and exploitation via an $\varepsilon$-\emph{greedy policy} \citep{Watkins1992}.

This approach directly yields a control law $\pi(s) = \arg\max_{a} Q_t(s,a)$ from experiences, but a tabular $Q$ scales as $\mathcal{O}(|S||\mathcal{A}|)$ and cannot generalize across large, continuous, or high-dimensional descriptions of traffic states—typical in TSC.

\emph{Deep Q-learning (DQN)} \citep{Mnih2015Human} addresses this by approximating $Q(s,a)$ with a neural network $Q_{\theta}(s,a)$ and minimizing a temporal-difference regression loss over mini-batches drawn from an experience replay memory $\mathcal{D}$:

{\setlength{\mathindent}{0pt}
\begin{equation}
\begin{aligned}
&\text{\ \ \ \ \ \ \ \ \ \ }\mathcal{L}(\theta) = \mathbb{E}_{(s,a,r,s')\sim\mathcal{D}}\!\left[\big(y - Q_{\theta}(s,a)\big)^2\right]\\
&\text{\ \ \ \ \ \ \ \ \ \ }y = r + \gamma \max_{a'\in \mathcal{A}} Q_{\bar{\theta}}(s',a')
\end{aligned}
\end{equation}
}

\noindent
Here, $Q_{\theta}$, represents the Q network and is a \emph{trainable} parametric approximator whose weights $\theta$ are updated via stochastic gradient steps to minimize $\mathcal{L}(\theta)$; $\mathcal{D}$ stores past \emph{experiences} to decorrelate samples; $\mathcal{L}(\theta)$ is the mean-squared TD loss; $Q_{\bar{\theta}}$ is a slowly updated \emph{target network} which is a lagged copy of $Q_{\theta}$, used solely to compute the bootstrapped target \(y\); the \emph{online network} \(Q_{\theta}\) produces the prediction being trained. In practice, its parameters are copied from the online network at fixed intervals ($\bar{\theta} \leftarrow \theta$ every $C$ training steps). Finally, $y$ is the one-step TD \emph{target} computed from the next observation $s'$ using this fixed snapshot $Q_{\bar{\theta}}$.

\noindent
To reduce the overestimation bias of the $\max$ operator, \emph{Double DQN (DDQN)} \citep{vanHasselt2016Deep} decouples action selection and evaluation in the target:
\begin{equation}
y_{\mathrm{DDQN}} = r + \gamma \, Q_{\bar{\theta}}\!\Big(s', \arg\max_{a'\in \mathcal{A}} Q_{\theta}(s',a')\Big).
\end{equation}
\noindent
In this target, the next action is selected by the online network $Q_{\theta}$ but evaluated by the target network $Q_{\bar{\theta}}$, which mitigates the positive bias.

With these components—function approximation for scalability, replay memory for decorrelation, and target networks (plus DDQN) for bias/variance control—value-based DRL retains the experience-driven character of Q-learning while making it practical for the rich state representations encountered in traffic signal control \citep{Rasheed2020}.
\subsection{Region-based MARL for Traffic Signal Control}
\label{sec:region-based-marl-for-tsc}

\hspace*{1em}We model urban traffic signal control as a multi-agent decision process in which multiple intersections act concurrently and are organized into regions that capture tightly interdependent behavior. Here, we provide this formulation to fix the core entities, notation that will carry through the rest of the paper.

\noindent\textbf{Regional Agent.}
In our formulation we distinguish the learning module from the physical signal controller. A \emph{regional agent} refers to the decision-making model parameterized by a DDQN, as introduced in \S\ref{sec:rl-based-modeling-of-tsc-problem}. This DDQN represents a policy that maps an observed state to action values. In other words, the agent is the learner: it is the function approximator that is optimized during training.

\begin{figure}[t]
    \centering

    \begin{subfigure}[b]{0.45\textwidth}
        \centering
        \includegraphics[width=0.38\textwidth, height=0.38\textwidth, keepaspectratio]{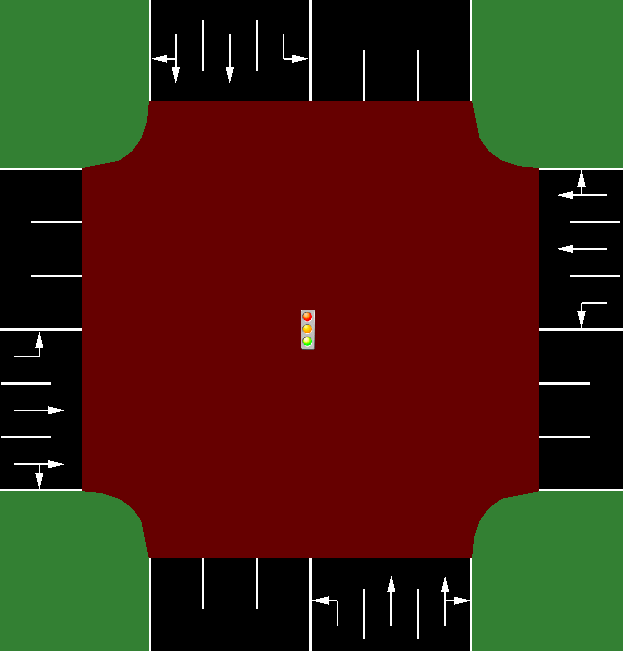}
        \caption{Cross-Intersection}
        \label{fig:crossroads}
    \end{subfigure}
    \hfill
    \begin{minipage}[b]{0.45\textwidth}

        \centering

        \begin{subfigure}[b]{0.38\textwidth}
            \centering
            \includegraphics[width=\textwidth, height=\textwidth, keepaspectratio]{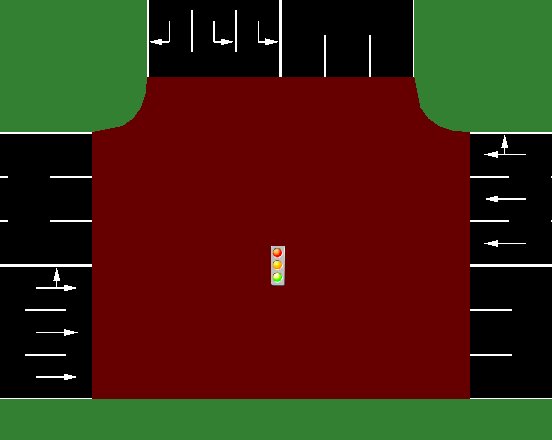}
            \caption{south blocked}
            \label{fig:t_south}
        \end{subfigure}
        \hfill
        \begin{subfigure}[b]{0.38\textwidth}
            \centering
            \includegraphics[width=\textwidth, height=\textwidth, keepaspectratio]{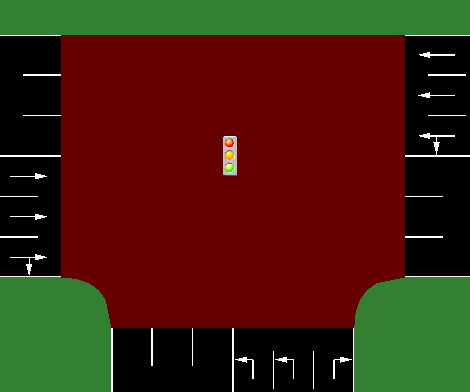}
            \caption{north blocked}
            \label{fig:t_north}
        \end{subfigure}

        \vspace{0.5em}

        \begin{subfigure}[b]{0.38\textwidth}
            \centering
            \includegraphics[width=\textwidth, height=\textwidth, keepaspectratio]{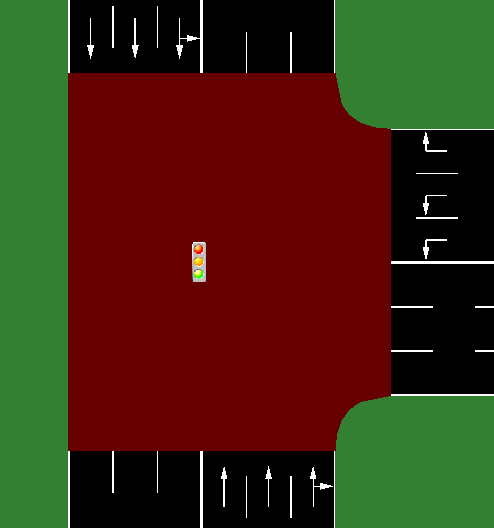}
            \caption{west blocked}
            \label{fig:t_west}

        \end{subfigure}
        \hfill
        \begin{subfigure}[b]{0.38\textwidth}
            \centering
            \includegraphics[width=\textwidth, height=\textwidth, keepaspectratio]{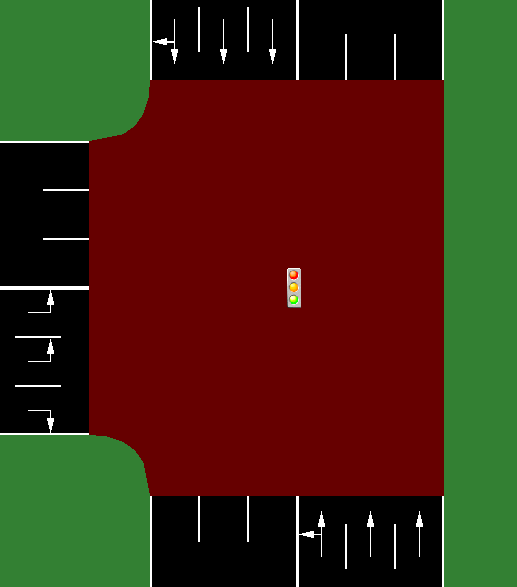}
            \caption{east blocked}
              \label{fig:t_east}
        \end{subfigure}

        \label{fig:tvariants}
    \end{minipage}

    \caption{Representative intersection topologies considered. 
    \textbf{(a)}: Cross-Intersection with all approaches present. 
    \textbf{(b-e)}: Different geometries of T-Intersections in the network.}
    \label{fig:intersection-topologies}
\end{figure}

\noindent\textbf{Intersection Agent.}
An \emph{intersection agent} is an individual signalized intersection that executes control within the traffic environment. At each control step, the intersection agent observes its own traffic conditions, encodes them as a state vector, and queries the DDQN policy associated with it to obtain and perform the next admissible signal operation. Intersection agents are heterogeneous. To better reflect real urban networks, we include both cross-intersection and T-intersection with their variants, which are depicted in Fig.~\ref{fig:intersection-topologies}; these geometries admit different non-conflicting movements and therefore expose different admissible actions. This geometric structure also influences how features are presented in their state representation when it queries a regional agent. As a result, multiple distinct intersection agents (with different layouts and feasible phase sets) can be served by the same regional agent, while still behaving in a coordinated manner.

\noindent\textbf{Region.}
A \emph{region} ${R}$ is a set of intersection agents whose behaviors are strongly interconnected: queues, spillback, and discharge at one intersection agent can immediately influence its neighbors. In practice, such regions can be derived from high-density corridors, spatial adjacency, or any structural pattern that creates tight mutual dependence. A region is treated as a coordination unit during training, while execution at runtime remains intersection-level. We denote by $\mathcal{R}(i)$ the region-assignment mapping that returns the unique region to which intersection agent $i$ belongs to. The entire traffic network is partitioned into $K$ disjoint regions:

\begin{equation}
R_j \subseteq I,
\quad 
 I = \bigcup_{k=1}^{K} R_k, 
\qquad 
R_j \cap R_\ell = \varnothing \ \text{for}\ j \neq \ell
\label{eq:inrersection-agenti-Rdef}
\end{equation}

Here, $I$ denotes the set of all intersection agents in the network.

\begin{figure}[h] 
    \centering
    \includegraphics[width=1.0\linewidth]{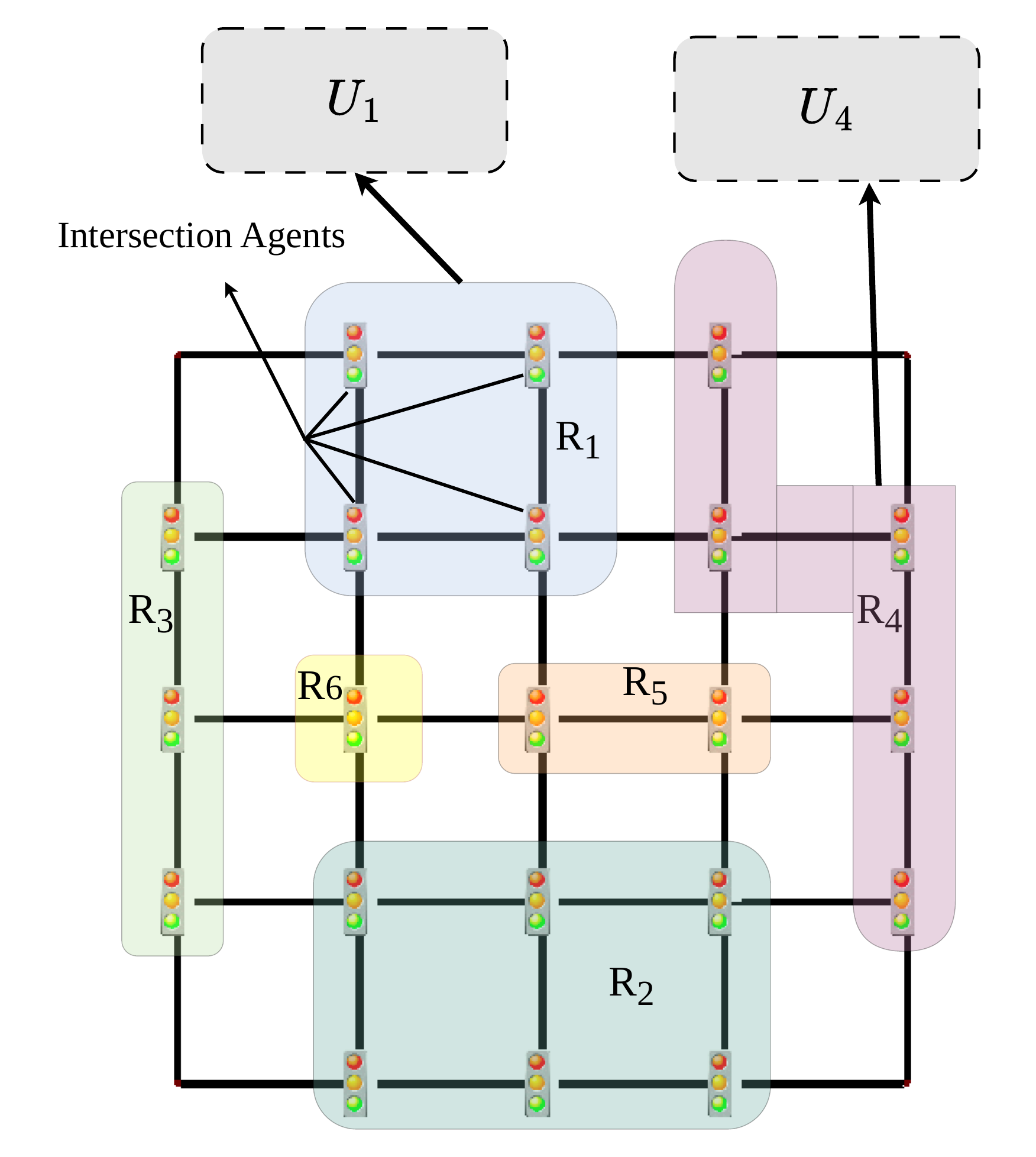}
    \caption{Region-based MARL for TSC}
    \label{fig:problemdef}
\end{figure}

Let $U$ denote the set of all regional agent sets. Each region ${R}_k$ is associated with a regional agent set $U_k \in {U}$, with $|U_k| \ge 1$ \eqref{eq:U-def}. $U_k$ contains a single shared agent or multiple regional agents. At runtime, each intersection agent $i \in {R}_k$ queries a designated regional agent $u_i \in U_k$ for control, where $u_i$ denotes the regional agent assigned to coordinate control for intersection $i$. The general problem formulation in this setting is depicted in Fig. \ref{fig:problemdef}.
\begin{equation}
 U = \bigcup_{k=1}^{K} U_k, 
\qquad U_k \neq \varnothing 
\label{eq:U-def}
\end{equation}


\noindent\textbf{Action Space.}
At each decision step $n$, the intersection agent $i$ executes an action $a_i^{(n)}$ that selects which signal phase to activate and for how long it will remain active. For a standard cross-intersection, we define four admissible logics: \texttt{NS\_S} (north--south through movements only), \texttt{NS\_L} (north--south protected left turns), \texttt{EW\_S} (east--west through movements only), and \texttt{EW\_L} (east--west protected left turns). Each logic can be executed with one of two discrete duration options: a \emph{short} option of $g_s$ seconds or a \emph{long} option of $g_l$ seconds. This yields eight concrete phases (\texttt{NS\_S short}, \texttt{NS\_S long}, \texttt{NS\_L short}, \texttt{NS\_L long}, \texttt{EW\_S short}, \texttt{EW\_S long}, \texttt{EW\_L short}, \texttt{EW\_L long}) that the policy may select at runtime. The inclusion of multiple duration choices allows the controller to modulate timing granularity without assuming a fixed control interval, making the action space closer to realistic actuated signal practice. 

\begin{figure}[t]
    \centering
    \includegraphics[width=1.0\linewidth]{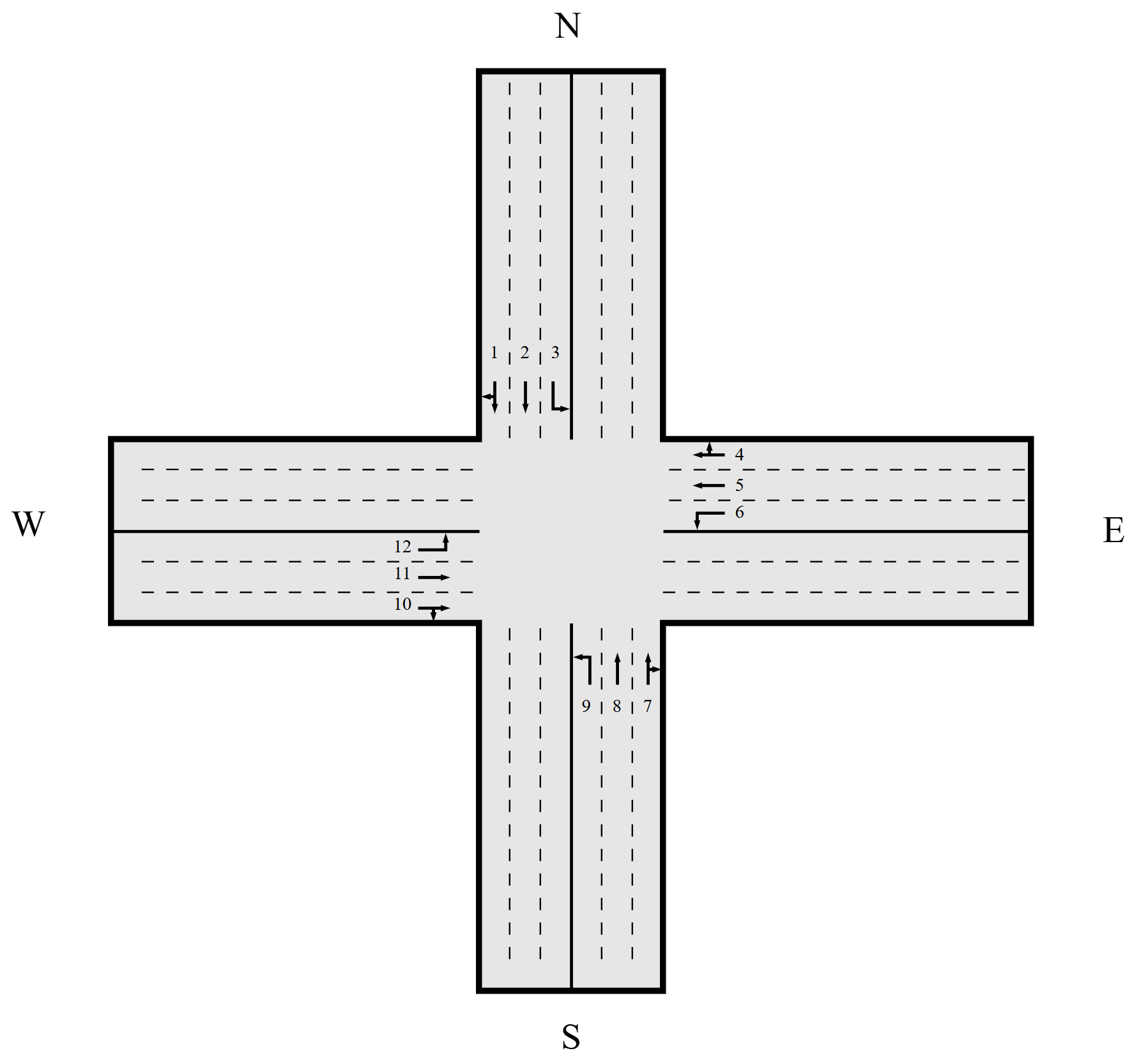}
    \caption{Base geometry and lane indices used to define phases at a standard cross-intersection (cf.\ Table~\ref{tab:phase-lanes}).}
    \label{fig:lane-indexing}
\end{figure}

When switching between different phases, a yellow/all-red clearance of 3\,s is enforced for safety. Transitions that simply continue the same logic at a different duration do not require an intermediate clearance. For T-intersections, which lack one approach of the crossroads, the action space $\mathcal{A}_i$
 is defined as the feasible subset of these same phases. In other words, each intersection agent’s admissible action set $\mathcal{A}_i$ is intersection agent-specific and physically valid by construction. We formally define $\mathcal{A}^G$ as the global action space and, for each intersection $i$, let $\mathcal{A}_i \subseteq \mathcal{A}^G$ denote its admissible (topology-valid) action set.
Fig.~\ref{fig:lane-indexing} depicts the base geometry and lane indices for a standard four-leg intersection, and Table~\ref{tab:phase-lanes} summarizes the global action space, $\mathcal{A}^G$.

\begin{table}[t]
    \centering
    \caption{Global action space $\mathcal{A}^G$ for a standard cross-intersection, listing allowed movements per lane. Lanes marked (r) are right-turn only.}
    \label{tab:phase-lanes}
    \begin{tabular}{ll}
        \hline
        \textbf{Phase} & \textbf{Served lanes} \\
        \hline
        \texttt{NS\_S short, NS\_S long} & 1, 2, 7, 8 \\
        \texttt{NS\_L short, NS\_L long} & 1(r), 3, 7(r), 9 \\
        \texttt{EW\_S short, EW\_S long} & 4, 5, 10, 11 \\
        \texttt{EW\_L short, EW\_L long} & 4(r), 6, 10(r), 12 \\
        \hline
    \end{tabular}
\end{table}

Since each logic can be executed with two duration options, we adopt an event-driven schedule. After executing action $a_i(t_i^{(n)})$, the next action occurs when the executed green expires plus any required inter-phase clearance. Where $n \in \mathbb{N}$ indexes the decision steps at intersection agent $i$. Formally:
\begin{equation}
    t_i^{(n+1)} \;=\; t_i^{(n)} 
    \;+\; g\!\big(a_i^{(n)}\big) 
    \;+\; c\!\big(a_i^{(n-1)}, a_i^{(n)}\big).
    \label{eq:event-driven}
\end{equation}
where $g(\cdot)$ denotes the selected action's green duration and 
$c(\cdot,\cdot)$ encodes whether a yellow/all-red clearance is needed when switching the phase.

\noindent\textbf{State.}
At each decision time $t$, intersection agent $i$ aggregates its observations into a fixed-length state vector $s_i^{(t)}\in S_i\subseteq\mathbb{R}^d$, where $S_i$ denotes the state space. This vector is a compact collection of informative traffic-flow features around intersection $i$. The state $s_i^{(t)}$ is then provided to the designated regional agent for $\mathcal{R}(i)$, which evaluates the admissible actions in $\mathcal{A}_i$ and returns a control decision. During learning, these states serve as the inputs on which the regional agent set is optimized to favor actions that improve long-term network performance. 

\noindent\textbf{Reward.}
At each decision time $t_i^{(n)}$, we attribute to intersection agent $i$ a scalar reward $r_i(t_i^{(n)})\in\mathbb{R}$ that reflects the performance delivered at $i$ and $\mathcal{R}(i)$. Because the outcomes of an executed action take effect with latency, this evaluation is computed after a delay and assigned back to the decision taken at $t$. Without committing to a specific shaping, $r_i(t_i^{(n)})$ should be read as an aggregation aligned with congestion mitigation and efficient discharge. The designated regional agent for $i$ uses these delayed rewards along with the observed states and chosen actions to improve decision making for long-horizon performance.

\section{Approach: Semi-Centralized Training, Decentralized Execution}

\hspace*{1em}In this section, we detail our proposed SEMI-CTDE architecture for ATSC.

\subsection{Region Formation}
\label{sec:region-formation}

\hspace*{1em}In the context of optimizing traffic signal control, the formation of regions serves as a critical strategy to enable effective coordination among intersection agents. The primary goal of region formation is to optimize traffic flow by grouping intersections that exhibit strong mutual dependence in their traffic dynamics. This approach is particularly valuable for managing the interdependencies between tightly coupled intersections, where congestion, queues, and spillback at one intersection can significantly affect its neighbors. Without coordinated control, these spatial and temporal interdependencies can result in suboptimal management, leading to increased delays and inefficiencies in traffic flow. By grouping intersections with similar traffic dynamics, regions partition the network into subsets of intersections requiring coordinated control policy.

We adopt the \emph{alpha-cut method} in fuzzy graphs, as proposed in \citet{9109698}, to partition the network (see Algorithm \ref{alg:region-formation}). The traffic network is modeled as a fuzzy graph, where each vertex represents an intersection and edges reflect dependencies based on traffic flow and congestion. The alpha-cut method partitions the graph into regions by selecting a threshold $\alpha$ that defines the strength of interdependence between intersections. The resulting regions, or ``correlated agent sets,'' consist of intersections that are tightly interdependent in terms of traffic dynamics, ensuring that coordination occurs primarily within these regions.

In SEMI-CTDE, regional agents oversee the behavior of intersection agents within their respective regions, so defining regions with these characteristics aligns naturally with this architecture. By facilitating coordinated actions within these regions, SEMI-CTDE builds on meaningful regional formation to improve traffic management and help to minimize delays. Intersection agents acting together within a region can more effectively mitigate congestion and prevent spillbacks, leading to better overall system performance compared to controlling the entire traffic network as a single entity.

\begin{algorithm}[t]
\caption{Region Formation via Fuzzy Graph $\alpha$-Cut}
\label{alg:region-formation}
\small
\begin{algorithmic}[1]
  \State \textbf{Inputs:} Intersection set $\mathcal{I}$; simulation horizon $n$; threshold $\alpha$.
  \State \textbf{Initialization:} Define directed graph $G=(V,E)$ with $V \gets \mathcal{I}$, $E \gets \emptyset$.

  \Statex \textit{Warm-up phase: collect congestion data}
  \For{$t = 1$ to $n$}
    \ForAll{intersection $i \in \mathcal{I}$}
      \State Update cumulative queues on incoming edges of $i$.
    \EndFor
  \EndFor
  \ForAll{intersection $i \in \mathcal{I}$}
    \State $\text{queue}(i) \gets$ average queue length on incoming edges of $i$.
  \EndFor

  \Statex \textit{Graph construction: assign congestion weights to edges}
  \ForAll{intersection $i \in \mathcal{I}$}
    \ForAll{downstream intersection $j$ reachable from $i$}
      \State Add $(i,j)$ to $E$ with weight $c_{ij} \gets \text{queue}(i)$.

    \EndFor
  \EndFor

  \Statex \textit{Fuzzy memberships}
  \ForAll{vertex $v \in V$}
    \State $\sigma(v) \gets$ avg congestion on incoming edges of $v$.

  \EndFor
  \ForAll{edge $(u,v) \in E$}
    \State $p_{uv} \gets \frac{c(u,v)}{\sum_j c(j,v)}$.
    \State $\mu(u,v) \gets \min(\sigma(u), \sigma(v)) \cdot p_{uv}$.
  \EndFor

  \Statex \textit{Apply $\alpha$-cut on edges}
  \ForAll{edge $(u,v) \in E$}
    \If{$\mu(u,v) < \alpha$}
      \State Remove edge $(u,v)$ from $E$.
    \EndIf
  \EndFor

  \Statex \textit{Region extraction}
  \State Construct undirected graph $G_{\alpha}$ from $G$ by keeping only edges with $\mu(u,v) \ge \alpha$.
  \State Perform connected component analysis on $G_{\alpha}$.

  \State \textbf{Output:} Regions $\{R_1, \dots, R_K\}$, where each $R_k$ is a connected group of intersections.

\end{algorithmic}
\end{algorithm}

\subsection{State Space Design}
\label{sec:state-design}

\hspace*{1em}In MADRL based traffic signal control, state representation is crucial since the chosen action at each decision step is highly dependent on the perceived information from the local and regional environment. In order to better address the local and regional context in state space design, SEMI-CTDE requires each intersection agent  $i$ to expose composite $s_i$ to its regional agent. We conceptualize $s_i$ as a modular concatenation
\begin{equation}
\label{eq:state-build}
s_i \;=\; \big[\, s_i^{\mathrm{local}} \;\Vert\; s_i^{\mathrm{regional}} \,\big],
\end{equation}
\noindent where $s_i\in\mathbb{R}^{d}$ and $\Vert$ denotes concatenation.

\noindent The local component captures intersection-specific operating context and the regional component summarizes the surrounding coordination context within $\mathcal{R}(i)$. Because execution is decentralized, SEMI-CTDE tolerates richer state representations per intersection agent while retaining coordinated learning benefits.

In SEMI-CTDE, both components are instantiated as a concatenation of feature sub-vectors with a fixed goal. In the following, we define $s_i^{\mathrm{local}}$ and $s_i^{\mathrm{regional}}$.

\noindent\textbf{Local State $s_i^{\mathrm{local}}$:}
The local state is limited to intersection agent $i$ and its immediate approaches, encoding only information that can be measured from its own incoming/outgoing movements and fixed topological properties. Its role is to describe how $i$ is currently operating or has operated in the very recent past and what structural constraints shape feasible control patterns. Consequently, we instantiate $s_i^{\mathrm{local}}$ using three main features:

\begin{equation}
    s_i^{\mathrm{local}}
    \;=\;
    \big[
        \phi_i^{p}
        \;\Vert\;
        \phi_i^{t}
        \;\Vert\;
        \phi_i^{s}
    \big]\quad.
\label{eq:local-state}
\end{equation}

\noindent{Local phase context $\phi_i^{{p}}$:} This feature characterizes the current control strategy at intersection agent $i$, providing the agent with short-term operational memory of how the signal is behaving. Typical elements include the current action or active phase, an indicator of whether the intersection is in an intermediate state such as a yellow transition, and the number of recent logical phase switches over a short horizon. These features help the agent reason about phase continuity, switching penalties and existence of stop-and-go behavior when selecting the next action.

\noindent{Local throughput features $\phi_i^{{t}}$:}
This feature summarizes how effectively intersection $i$ is processing vehicles on its immediate approaches. It may include lane- or approach-level queue lengths, average speeds, cumulative or mean waiting times, local demand or arrival rates, and simple discharge or outflow measures on the incoming and outgoing links. By exposing both congestion indicators and throughput measures, this component allows the policy to directly target reduced delay, controlled queue growth, and efficient utilization of green time at the local scale.

\noindent{Local spatial features $\phi_i^{{s}}$:}
This feature encodes invariant structural properties of intersection $i$ that shape its role in the network. Examples include whether intersection $i$ is a T or cross-intersection, its connectivity pattern, such as direction of incoming approaches with respect to itself, and a role indicator such as perimeter vs.\ interior intersection in grid-like networks. Such features enable the learned policy to associate consistent behaviors with similar geometric and topological roles, improving generalization across heterogeneous intersections while preserving a uniform local-state encoding.

\noindent\textbf{Regional State $s_i^{\mathrm{regional}}$:}
The regional component equips the intersection agent with a top-down view that extends beyond the immediate approaches to intersection $i$ and summarizes how its surrounding intersections in $\mathcal{R}(i)$ operate. The exact same goals in Local State $s_i^{\mathrm{local}}$ are followed here, and the distinction is in the scope of the encoded features. Accordingly, we define $s_i^{\mathrm{regional}}$ as below:

\begin{equation}
    s_i^{\mathrm{regional}}
    \;=\;
    \big[
        \psi_i^{p}
        \;\Vert\;
        \psi_i^{t}
        \;\Vert\;
        \psi_i^{s}
    \big].
\label{eq:regional-state}
\end{equation}

\noindent A key design choice is the receptive field: one may use (a) whole-region summaries that pool information over all intersection agents in $\mathcal{R}(i)$ to expose global imbalances and overall loading, or (b) bounded-neighborhood summaries that emphasize the most influential vicinity to sharpen local coordination. Either option is compatible with SEMI-CTDE, and the choice should reflect region size and complexity of learning network in terms of number of learnable parameters and state space size. To convert a large number of regional features into a fixed-length representation, one can employ invariant pooling (e.g., sum/mean/max over sets) or directionally organized summaries to encode directional imbalance where appropriate.

\noindent{Regional phase context $\psi_i^{p}$:}
 Instead of only the current phase of a single intersection, this feature encodes aggregated or neighborhood-level information about active phases, recent switching behavior, or synchronization patterns within $\mathcal{R}(i)$. This enables the intersection agent to infer whether it operates within a mostly compatible or mostly conflicting regional signal configuration when selecting its next action.

\noindent{Regional throughput features $\psi_i^{t}$:}
Regional throughput summarizes congestion and discharge indicators at the regional scale, constructed via the receptive field choices outlined earlier. It aggregates quantities such as regional queue burdens, delays, or utilization over $\mathcal{R}(i)$, while preserving a fixed-size encoding. In addition, SEMI-CTDE recommends explicitly incorporating throughput indicators at region boundaries, so that the regional signal captures how well the region exchanges traffic with its surroundings and where coordinated discharge is most needed.

\noindent{Regional spatial features $\psi_i^{s}$:}
This feature encodes structural information at the regional scale in a way that remains portable across alternative partitions. Typical elements include the relative position of intersection $i$ within $\mathcal{R}(i)$, descriptors of the region’s layout, and explicit boundary indicators for nodes or approaches adjacent to other regions. These features allow the policy to associate distinct coordination behaviors with different regional roles and to learn about cross-region interactions.

When instantiating concrete features within each local and regional component, two guidelines should be followed. First, both local and regional components should expose features that are closely aligned with the evaluation objectives. This enables improvements in the learned policy to translate directly into gains in the chosen performance metrics. Second, scale-comparable encodings should be maintained to normalize the number of approaches for each intersection and region size. This helps to keep shared policies comparable across intersections and regions of varying topologies and sizes. Collectively, this modular state design produces a compact and interpretable representation that distinguishes local from regional context, facilitates decentralized execution, and enables centralized learning within each region via shared regional agents.

\subsection{Reward Function Design}
\label{sec:reward-design}
In traffic-signal control, the general network-level objective is to minimize average travel time across the network. Because this global performance objective is neither directly observable at decision time nor immediately attributable to individual control actions, reward formulations instead rely on measurable quantities that approximate or correlate with travel time. In SEMI-CTDE, each intersection agent $i$, between two consecutive decision times $[\,t_i^{(n)},\, t_i^{(n+1)}\,)$, receives a composite reward attributed to the action taken at $t_i^{(n)}$:
\begin{equation}
    r_i \;=\; \beta_{\mathcal{R}(i)}\, r_i^{\mathrm{local}}
    \;+\; (1 - \beta_{\mathcal{R}(i)})\, r_i^{\mathrm{regional}}\! \quad.
    \label{eq:reward-decomp}
\end{equation}

Here, the local term evaluates traffic flow quality at intersection $i$, and the regional term captures coordination effects within region $\mathcal{R}(i)$. The weights $(\beta_{\mathcal{R}(i)}, 1-\beta_{\mathcal{R}(i)})$ are defined per region and determine the trade-off between local and regional objectives. These weights are kept constant to promote training stability and are chosen empirically. These two terms should be designed with careful attention to several practical considerations; we summarize the key guidelines below.

\noindent\textbf{Local reward $r_i^{\mathrm{local}}$:}
At each decision step, this term assesses the quality of traffic flow performance at intersection $i$ by considering reductions in vehicle delay and control of queue growth, thereby promoting discharge and spillback avoidance. To keep comparisons fair across heterogeneous geometries, contributing quantities should be normalized with respect to physical geometry. The features used should be jointly designed with the local state vector in \S\ref{sec:state-design}, so that what the agent perceives is tightly aligned with what it is incentivized to improve. To mitigate stop-and-go behavior at intersections, it is also helpful to penalize frequent phase switches which could introduce lost time and inefficiencies. A design including some or all these elements can better approximate the general travel time objective.

\noindent\textbf{Regional reward $r_i^{\mathrm{regional}}$:}
This term evaluates performance across the entire region $\mathcal{R}(i)$, encouraging coordination that addresses shared bottlenecks and improves region-boundary throughput. It should retain the same core design principles as the local term, but extended to the regional scale. These principles include favoring region-level delay and queue mitigation, employing geometry-based normalization, considering the regional state vector in \S\ref{sec:state-design} and penalizing frequent phase switches. For the regional reward term, like receptive field choice in regional state design, two design choices are available. Whole-region summaries that pool information over neighborhood intersection agents in $\mathcal{R}(i)$ to expose global imbalances and overall loading, and bounded-neighborhood summaries that emphasize the most influential vicinity to sharpen local coordination. Either option is compatible with SEMI-CTDE, and the selection should reflect region size, topology, and corridor structure.

Overall, the composite reward in Eq. \eqref{eq:reward-decomp} provides feedback that reflects both local and regional coordination objectives. This supports consistent reward assignment under SEMI-CTDE's decentralized execution paradigm.
In addition, Region-specific weights  $(\beta, 1-\beta)$ allow each region to calibrate the trade-off between local performance and regional coordination. Each region can adapt to its own topology and demand pattern while maintaining the consistency of the overall architecture.

\subsection{Proposed Architecture (SEMI\texorpdfstring{-}{-}CTDE)}
\label{sec:semi-ctde}

\begin{figure*}[t]
    \centering
\includegraphics[width=\textwidth]{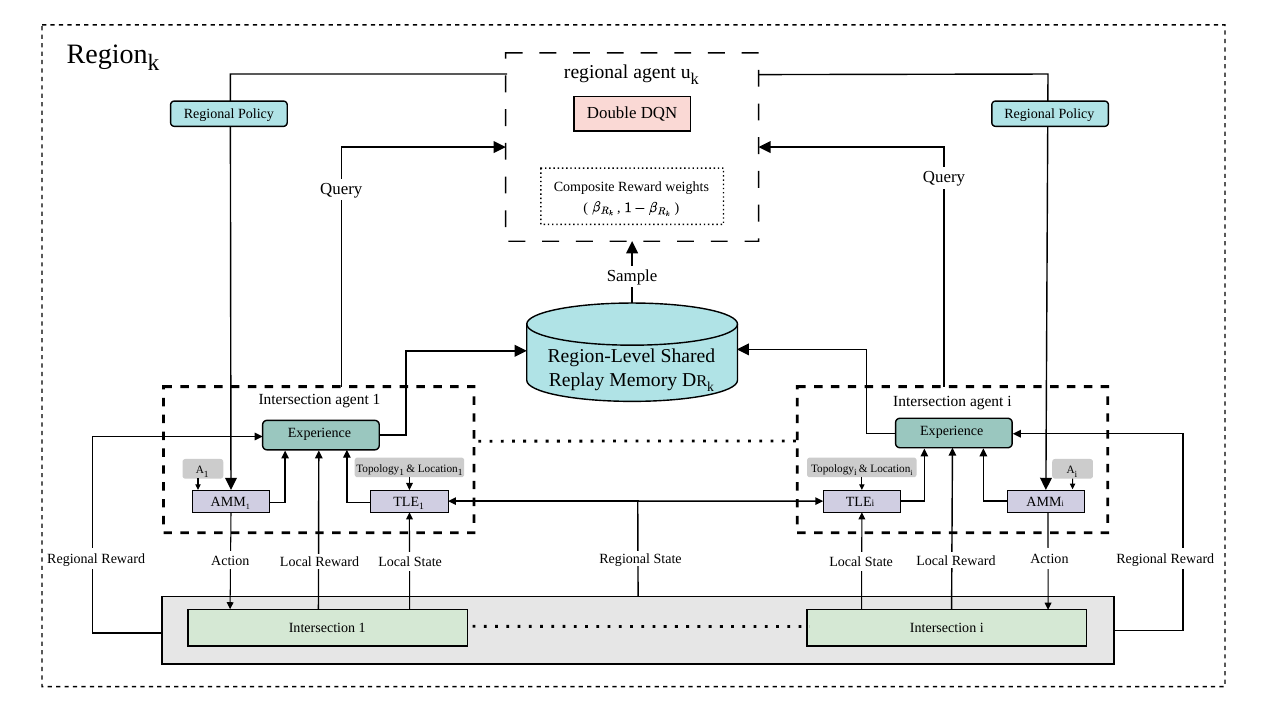}
    \caption{SEMI-CTDE architecture inside each region.}
    \label{fig:semi-ctde-architecture}
\end{figure*}

\hspace*{1em}Given a fixed partition of the network into regions \S\ref{sec:region-formation}, we now describe how the \emph{Semi-Centralized Training, Decentralized Execution} paradigm operates over these regions. Fig.~\ref{fig:semi-ctde-architecture} depicts the proposed architecture of SEMI-CTDE inside each region.

\subsubsection{Instantiation of Region Entities}
\label{sec:region-instantiation}
\begin{figure}[t]
    \centering
    \includegraphics[width=\columnwidth]{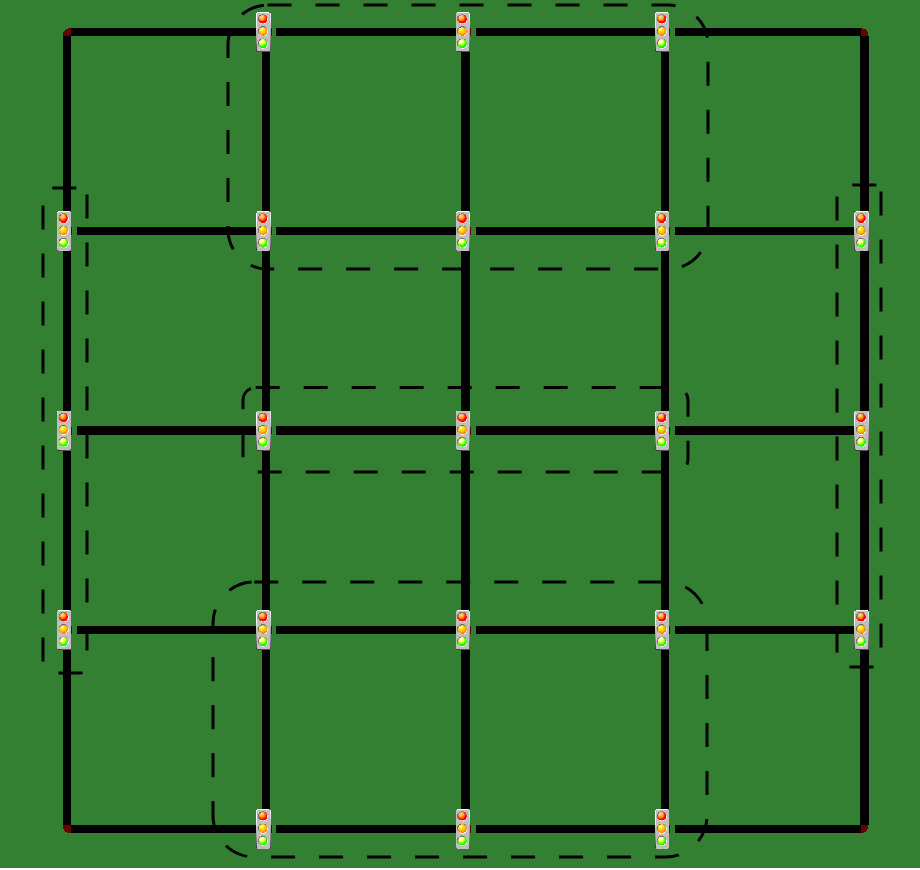}
    \caption{Network partition into regions $\{R_k\}_{k=1}^{K}$ used in this paper. 
    Traffic lights denote intersection agents; dashed lines mark region boundaries. 
    The partition is obtained via the method described in ~\S\ref{sec:region-formation}.}
    \label{fig:regioning-map}
\end{figure}

Let $\{R_1,\ldots,R_K\}$ denote disjoint regions over the set of intersection agents $I$ (Eq. \eqref{eq:inrersection-agenti-Rdef}). Each region $R_k$ is served by a \emph{single} regional agent $u_k$ parameterized by a DDQN. Thus, for each $R_k$, ${U}_k=\{u_k\}$. Every intersection agent $i\in R_k$ queries $u_k$ for control, so $u_i = u_k$. Intersection agents differ in geometry and feasible phase sets but share the same regional agent within $R_k$. The network partition employed throughout this work was obtained using the procedure of ~\S\ref{sec:region-formation} is illustrated in Fig.~\ref{fig:regioning-map}.

\subsubsection{Intersection Agent-Side Processing Modules}
\label{sec:intersetcion-agent-side-processing-modules}
Each intersection agent $i$ contains two \emph{independent, non-trainable} processing units that operate before and after querying its regional agent. These units depend on the agent's geometry and its location within the region but do not depend on the regional agent or its parameters. We refer to them as the \emph{Topology \& Location Encoder (TLE)} and the \emph{Action Mapping Module (AMM)}.

\medskip
\noindent(1) Topology \& Location Encoder (TLE): At each $t_i^{(n)}$, intersection agent $i$ encodes observations into a topology- and location-aware state vector $s_i^{\mathrm{enc}}$ using a routine that depends only on the intersection agent’s physical type and its placement within the region. This routine is independent of the regional agent. The TLE’s responsibility is orthogonal to the concrete state definition. Regardless of which features are chosen, location and topology always yield intersection agent-specific standardization. The TLE is where these effects are resolved so the regional agent receives a region-comparable, fixed-layout encoding. Formally, for each intersection agent $i$ let $\mathrm{TLE}_i$ denote the encoder. At each $t_i^{(n)}$:
\begin{equation}
\label{eq:tle-encode}
s_i^{\mathrm{enc}} \;=\; \mathrm{TLE}_i\!\big(s_i\big).
\end{equation}

\medskip
\noindent(2) Action Mapping Module (AMM):
Upon receiving the global $Q$ vector from the regional agent, the AMM reconciles these scores with the intersection agent’s heterogeneous, geometry-specific action space. It applies a fixed, topology-aware mapping from the global action space $\mathcal{A}^G$ to the intersection agent’s admissible set $\mathcal{A}_i$, discarding ineligible entries so selection (and any exploration) occurs only over valid actions. Crucially, when writing the transition to the regional replay memory, the AMM stores the intersection-valid, actually executed $Q$ value corresponding to the selected admissible action after mapping. This module is rule-based and independent of the regional agent. Its role is to enforce physical validity in both decision-making and experience gathering. Formally, for each intersection agent $i$ define a deterministic mapping 
$\mathrm{AMM}_i$:
\begin{equation}
\label{eq:amm-extract}
Q_i(t_i^{(n)}) \;=\; \mathrm{AMM}_i\!\left(Q_{\theta_k}\!\big(s_i^{\mathrm{enc}}(t_i^{(n)}),\cdot\big)\right).
\end{equation}

\noindent
Together, TLE (before query) and AMM (after query) let heterogeneous intersection agents expose consistent states to a shared regional agent and execute physically valid controls from global $Q$ values. This design preserves decentralized execution while enabling semi-centralized learning under shared regional parameterization.

\subsubsection{Intersection Agents and the Regional Agent}
\label{sec:intersetcion-agents-and-regional-agents}
Within each $R_k$, the regional agent $u_k$ serves as a shared global controller for all intersection agents $i\in R_k$. The regional agent is a single DDQN with parameters $\theta_k$ and $\bar{\theta}_k$  and a region-level replay memory $D_{R_k}$. Intersection agents act as local controllers: they (i) acquire raw sensor signals locally, (ii) run the TLE to produce a fixed-layout, region-comparable state, (iii) query $u_k$ for $Q$-values, and (iv) invoke their AMM to align the global $Q$ with their topology-specific action set $\mathcal{A}_i$, (v) execute the control signal, and store the intersection agent-valid executed action in the experience written to $D_{R_k}$.

\subsubsection{Asynchronous experience gathering and querying}
\label{sec:async-interaction-and-experience-aggregation}
Intersections agents start an episode synchronously but diverge as soon as they commit to different phase durations, so decision step timing becomes intersection agent-specific and query times are therefore generally not aligned. Within each region, each intersection agent queries $u_k$ at intervals determined by Eq.~\eqref{eq:event-driven}, yielding distinct query times. Each intersection agent closes its preceding transition by observing the delayed reward attributable to $a^{(n)}_i$ and forming $(s,a,r,s')$. This event-driven mechanism accumulates experiences asynchronously yet maintains correct temporal feedback assignment, since rewards are attached to the action that occupied the green until its scheduled decision step.

\subsubsection{Regional Optimization and Parameter Sharing}
\label{sec:regional-optimization-and-parameter-sharing}
Within each $R_k$, a single regional agent $u_k$ maintains one parameter vector $\theta_k$ shared by all intersection agents in $R_k$. Experiences generated asynchronously by the intersection agents are written to the $D_{R_k}$. Updates to $\theta_k$ are computed from data sampled from $D_{R_k}$.  
This realizes semi-centralized training at region scope while preserving decentralized execution at the intersections. Parameter sharing ensures that all intersections in a region estimate values and select actions consistently. By storing experiences from all intersection agents in $D_{R_k}$, the regional agent learns from diverse intersection geometries, traffic conditions, and decision timing patterns, which promotes coordinated behavior.

\subsubsection{Execution-Time Decentralization}
\label{sec:execution-time-decenteralization}
After training, each intersection agent $i\in R_k$ acts locally by forming $s_i^{\mathrm{enc}}$, querying the frozen regional agent $u_k$, and performing the inferenced control signal. No inter-region communication is required at runtime. Coordination arises from regional parameter sharing during training and the state/reward alignment.

\subsubsection{Instantiation Degrees of Freedom}
\label{sec:instantiation-degrees-of-freedom}
SEMI-CTDE does not prescribe low-level training specifications. Beyond the state space (\S\ref{sec:state-design}) and reward function (\S\ref{sec:reward-design}) backbones, several design choices are intentionally left open to implementation. The \emph{target update rule} that governs synchronization between evaluation and target networks; the \emph{exploration policy} that converts admissible value estimates into behavior; the \emph{replay memory architecture} that specifies how experiences are stored and kept at region scope; and the \emph{batch construction policy} that determines how samples are drawn for learning steps. Keeping these degrees of freedom explicit makes the architecture broadly applicable, easy to port across learning algorithms, and well-suited to systematic ablations without altering the backbone.

We summarize the proposed SEMI-CTDE training algorithm in Algorithm \ref{alg:semi-ctde}.

\begin{algorithm}[t]
\caption{SEMI-CTDE}
\label{alg:semi-ctde}
\small
\begin{algorithmic}[1]
  \Statep{\textbf{Inputs:} Region partition $\{R_k\}_{k=1}^{K}$; global action space $\mathcal{A}$; reward weights $(\beta_{\mathcal{R}(i)}, 1-\beta_{\mathcal{R}(i)})$; green-duration function $g(\cdot)$; clearance time $c(\cdot,\cdot)$; exploration policy $\Pi_{\mathrm{explore}}$; training-step trigger $\mathcal{T}_{\mathrm{step}}$; target network update law $\mathcal{L}_{\mathrm{target}}$; minibatch sampler $\mathcal{B}_k:(D_{R_k},\,|B|)$; per-intersection agent $\mathrm{TLE}_i$; per-intersection agent $\mathrm{AMM}_i$.}
  \Statep{\textbf{Initialization:} For each region $k$: initialize regional agent parameters $\theta_k$,  $\bar{\theta}_k$ and shared replay $D_{R_k}$.}
  \For{episode $=1$ to $M$}
    \Statep{reset environment; $t\gets 0$}
    \While{episode not terminal}
      \ForAll{regions $k=1,\dots,K$}
        \ForAll{intersection agents $i\in{R}_k$ with $t=t_i^{(n)}$}
          \Statep{\textbf{State:} form $s_i$
           using ~\eqref{eq:state-build}}
          \Statep{\textbf{TLE:} gather $s_i^{\mathrm{enc}}(t_i^{(n)})$ using ~\eqref{eq:tle-encode}}
          \Statep{\textbf{AMM:} extract $Q_i(t)$ from $u_k$ using ~\eqref{eq:amm-extract}}
          \Statep{\textbf{Action selection:} $a_i^{(n)}\gets \Pi_{\mathrm{explore}}\big(Q_i(t)\big)$.}
          \Statep{\textbf{Actuation \& scheduling:} perform $a_i^{(n)}$ and schedule next query using ~\eqref{eq:event-driven}.}
        \EndFor
      \EndFor
      \Statep{advance simulation step to the next event time $t\gets \min_{i} t_i^{(n+1)}$.}
      \ForAll{regions $k=1,\dots,K$}
        \ForAll{intersections $i\in{R}_k$ with $t=t_i^{(n+1)}$}
          \Statep{observe next raw $s_i(t)$ and set $s_i^{\prime\,\mathrm{enc}}\gets \mathrm{TLE}_i\!\big(s_i(t)\big)$.}
          \Statep{\textbf{Rewards:} compute $r_i(t_i^{(n)})$ using ~\eqref{eq:reward-decomp}}
          \Statep{\textbf{Replay append:} push $\big(s_i^{\mathrm{enc}}(t_i^{(n)}),\, a_i^{(n)},\, r_i(t_i^{(n)}),\, s_i^{\prime\,\mathrm{enc}},\, \mathrm{done}_i;\, \big)$ to $D_{R_k}$.}
        \EndFor
        \If{$\mathcal{T}_{\mathrm{step}}$ triggers for region $k$ at time $t$}
          \State sample minibatch $B$ according to $\mathcal{B}_k$
          \State update $\theta_k$ using $B$.
          \If{$\mathcal{L}_{\mathrm{target}}$ triggers for region $k$}
            \State $\bar{\theta}_k \gets \mathcal{L}_{\mathrm{target}}(\theta_k, \bar{\theta}_k)$
          \EndIf
        \EndIf
      \EndFor
    \EndWhile
  \EndFor
\end{algorithmic}
\end{algorithm}

\section{Applying SEMI-CTDE: Two Implementations}

\hspace*{1em} In this section, we instantiate SEMI-CTDE with two concrete models that strictly follow the guidelines stated in \S\ref{sec:state-design} and \S\ref{sec:reward-design}. Both implementations adopt the same region partitioning, regional agent architecture, and interaction mechanics described in \S\ref{sec:semi-ctde}; they differ only in how the regional components of the state and reward are constructed.

We refer to these two models as \emph{RegionWide} and \emph{OneHop}. Their distinction stems from the choice of receptive field used in the regional state and reward: OneHop employs a bounded neighborhood-based receptive field, whereas RegionWide uses whole-region summaries. We first define the local state and reward used in the OneHop and RegionWide models and instantiate the AMM employed in them. We then detail the regional state and reward formulations
specific to each model.

\subsection{Local State and Reward Definitions}
\label{sec:local-components}

\hspace*{1em} Here we detail the identical  local state and reward formulations employed in OneHop and RegionWide models.
\subsubsection{Local State $s_i^{\mathrm{local}}$}
Consistent with the $s_i^{\mathrm{local}}$ definition in Eq.~\eqref{eq:local-state}, we instantiate the local state of intersection agent $i$ as the concatenation of $\phi_i^{t}, \phi_i^{p}$ and 
$\phi_i^{s}$.

\paragraph{$\phi_i^{p}$:}
To encode the current control strategy, we use current signal phase at $i$ represented by a one-hot vector
\begin{equation}
\phi_i^{p} = \text{phase}_i \in \{0,1\}^{\mathcal{A}^G}.
\end{equation}
where $\lvert \mathcal{A}^G \rvert = 8$ (see Table  \ref{tab:phase-lanes}).

\paragraph{$\phi_i^{t}$:}
To measure how effectively intersection $i$ is managing vehicles in immediate approaches, we group aggregating queue lengths caused by halted vehicles on incoming approaches into four scalars:
$q_i^{v}$, $q_i^{h}$, 
$q_i^{vl}$ 
and $q_i^{hl}$.
denoting the number of halted vehicles on vertical through approaches,  horizontal through approaches, and their associated left-turn lanes, respectively. These quantities are normalized over the immediate incoming approaches and left-turn lanes according to the intersection topology. We collect them as:
\begin{equation}
\phi_i^{t}
\;=\;
\big[
q_i^{v},
q_i^{h},
q_i^{vl},
q_i^{hl}
\big].
\end{equation}
We separate left-turn halted vehicles from their corresponding straight through movements to better align the state representation with the action space, where protected left-turn phases are controlled independently.

\paragraph{$\phi_i^{s}$:}
To encode structural property of intersections, we use a 5-dimensional one-hot vector:
\begin{equation}
\phi_i^{s} =
\begin{cases}
[1,0,0,0,0] & \text{west blocked},\\
[0,1,0,0,0] & \text{east blocked},\\
[0,0,1,0,0] & \text{south blocked},\\
[0,0,0,1,0] & \text{north blocked},\\
[0,0,0,0,1] & \text{4-way},
\end{cases}
\label{eq:type-onehot}
\end{equation}
indicating whether intersection $i$ is one of the four T-intersection variants or a 4-way intersection.

\subsubsection{Local Reward $r_i^{\mathrm{local}}$}
\label{sec:local-reward-exact}

Consistent with the design principles in \S\ref{sec:reward-design}, we instantiate the local reward at intersection $i$ as an average queue-length penalty over its immediate incoming approaches. Let $\bar{q}_i(n)$ denote the topology-normalized sum of queue lengths over all immediate approaches to intersection $i$ at decision step $n$. We define local reward at decision time $t_i^{(n)}$ as:
\begin{equation}
  r_i^{\mathrm{local}} \;=\;
   -\bar{q}_i(n),
\end{equation}

\noindent This choice directly penalizes congestion at intersection $i$ by assigning more negative rewards when more vehicles are halted,
by reusing exactly the queue lengths features that define $\phi_i^{t}$, this reward remains tightly aligned with what the agent perceives locally: actions that discharge queues and prevent spillback immediately reduce queue lengths at intersection $i$  and hence yield less negative local rewards.

\subsection{AMM Instantiation}
\label{sec:amm-instantiation}

\hspace*{1em} We instantiate the Action Mapping Module in both models using a topology-aware masking mechanism inspired by the admissibility masking mechanism proposed in~\citet{WANG2024104582}.

At each decision time $t_i^{(n)}$, the regional agent produces a global $Q$-value vector over the full action space $\mathcal{A}^G$. The AMM then applies an elementwise mask over this vector, setting the entries corresponding to actions not admissible in $\mathcal{A}_i$ to $-\infty$:
\begin{equation}
\mathrm{AMM}_i\!\left(Q_{\theta_k}\!\big(s_i^{\mathrm{enc}}(t_i^{(n)}),\cdot\big)\right)
=
\begin{cases}
Q_{\theta_k}(s_i^{\mathrm{enc}}(t_i^{(n)}), a), & a \in \mathcal{A}_i,\\[4pt]
-\infty, & a \notin \mathcal{A}_i,
\end{cases}
\qquad
\label{eq:amm-mask}
\end{equation}
This masking step guarantees that the subsequent $\arg\max$ (or $\epsilon$-greedy exploration) is performed strictly over the admissible subset $\mathcal{A}_i$, because all invalid actions are effectively removed from consideration.

\subsection{RegionWide model}
\label{sec:regionwide}
\begin{figure}[t]
    \centering
    \includegraphics[width=1.0\linewidth]{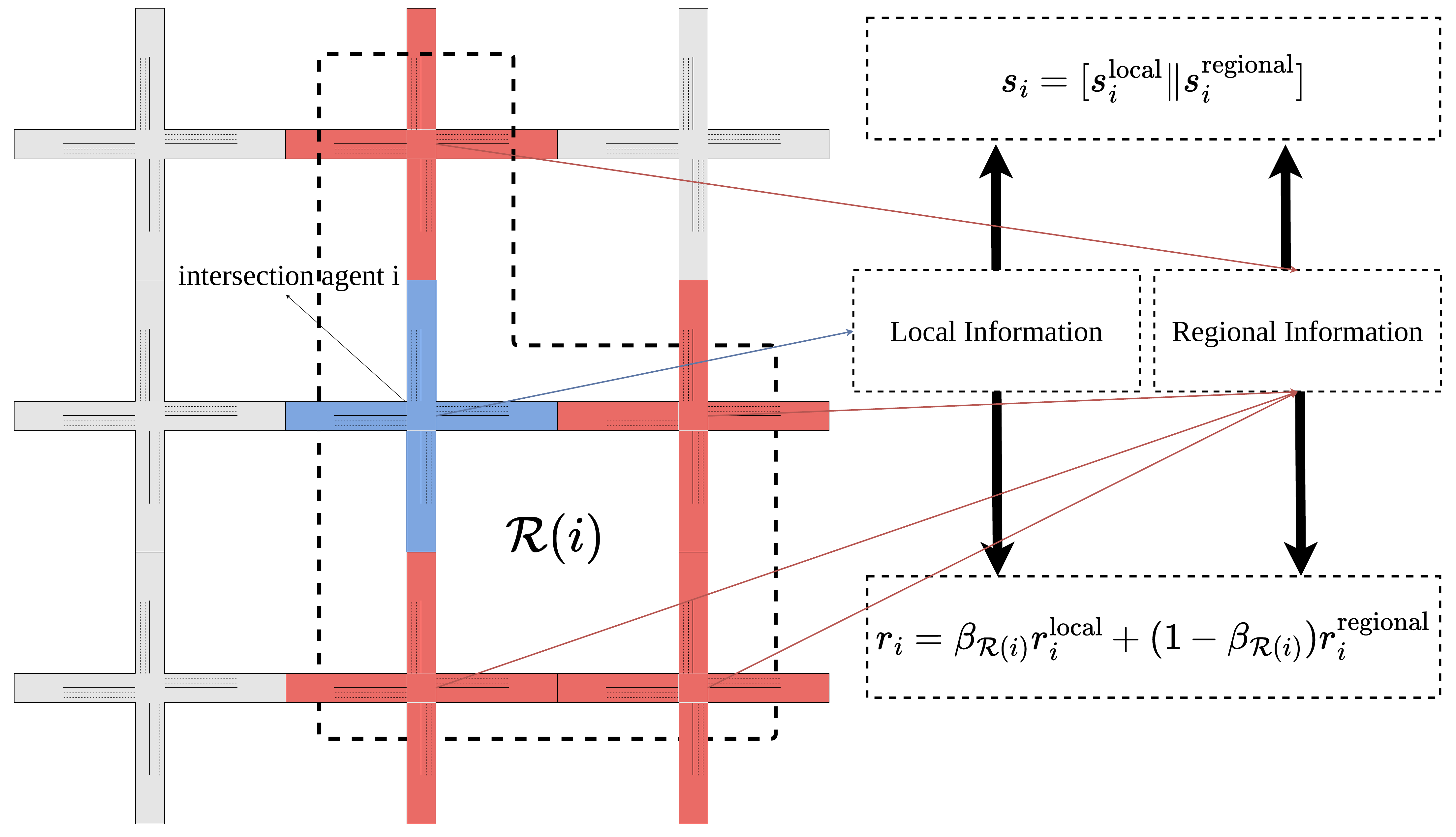}
    \caption{RegionWide model receptive field.}
    \label{fig:regionwide-rf}
\end{figure}

\hspace*{1em}RegionWide employs whole-region summaries receptive field to encode regional information. This scheme is presented in Fig. \ref{fig:regionwide-rf}.

\noindent\text{RegionWide Regional State $s_i^{\mathrm{regional}}$:}
In the RegionWide model, each intersection agent $i$ exposes to its regional agent a whole-region summary that aggregates information across all intersections within $\mathcal{R}(i)$. Consistent with the structured decomposition of $s_i^{\mathrm{regional}}$ in Eq.~\eqref{eq:regional-state}
we instantiate the regional features as follows.

\medskip
\noindent\emph{$\psi_i^{p}$:}
To represent the region-level phase configuration, we compute the proportion of intersections in $\mathcal{R}(i)$ currently executing each admissible logic in the global action set $\mathcal{A}^G$. Since $ \mathcal{A}^G$ contains four admissible logics, in this setup we obtain a four-dimensional vector
\begin{equation}
\psi_i^{p}
=
\frac{1}{\lvert \mathcal{R}(i) \rvert}
\left[
\sum_{j \in \mathcal{R}(i)} \mathbf{1}\{ \text{phase}_j = a_k \}
\right]_{k=1}^{4}.
\end{equation}

where $a_1,\ldots,a_4$ enumerate the admissible logics. This feature exposes the regional distribution of active logics, allowing the regional agent to infer global synchronization patterns within the region.

\medskip
\noindent\emph{$\psi_i^{t}$:}
The throughput block consists of three components capturing congestion, spillback, and regional exchange flows.

\smallskip
(i) 
We mirror the throughput structure used in the local state by incorporating
the neighbor’s halted-vehicle aggregates normalized according to intersection agent $j$'s topology
We aggregate these quantities across $\mathcal{R}(i)$ using regional averaging:
\begin{equation}
\bar{q}_{\mathcal{R}(i)}
=
\frac{1}{\lvert \mathcal{R}(i) \rvert}
\sum_{j \in \mathcal{R}(i)}
\big[
q_j^{v},
q_j^{h},
q_j^{vl},
q_j^{hl}
\big].
\end{equation}

\smallskip
(ii) Let each approach in the region have halted vehicles $q_a$, and let the spillback threshold be $\tau=15$, determined from approach length and width and vehicle size. The spillback ratio is:
\begin{equation}
\begin{gathered}
\text{IsCongested}(a)=
\begin{cases}
1 & \text{if } q_a>\tau \\
0 & \text{otherwise}
\end{cases}
\\[0.8em]
\chi_{R(i)}=
\frac{\sum_{a \in \mathcal{E}_{R(i)}} \text{IsCongested}(a)}{|\mathcal{E}_{R(i)}|}
\end{gathered}
\end{equation}

representing the fraction of approaches experiencing severe congestion.

\smallskip
(iii)
Let $\mathcal{E}_{\mathcal{R}(i)}^{\mathrm{boundary}}$ denote the set of boundary approaches, i.e., approaches whose upstream and downstream intersections belong to different regions. For each such approach $a$, let $\Delta{N}_a$ denote the change in the number of vehicles present between two consecutive regional state update intervals. Over an update duration of length $\Delta t$, we compute
\begin{equation}
f_{R(i)}
=
\frac{
  \sum_{a \in \mathcal{E}_{R(i)}^{\mathrm{boundary}}} \Delta N_a
}{
  \Delta t
},
\end{equation}

which provides an estimate of the region’s net exchange flow with its surroundings.

\smallskip
Collecting these three components yields the throughput feature:
\[
\psi_i^{t}
=
\big[
\bar{q}_{\mathcal{R}(i)},
\;\chi_{\mathcal{R}(i)},
\;f_{\mathcal{R}(i)}
\big].
\]

\medskip
\noindent\emph{$\psi_i^{s}$:}
To encode the relative placement of intersection $i$ within its region, we use a
four-dimensional binary vector that indicates, for each cardinal direction,
whether the corresponding one-hop neighbor lies outside $\mathcal{R}(i)$. Using
$\mathcal{N}_i^{\mathrm{hop}}$ to denote the directional neighbor set, we define
$\mathcal{N}_i^{\mathrm{hop}}$ to denote the directional neighbor set. We define:
\begin{equation}
\psi_i^{s}
=
\big[
\mathbf{1}\{\, j \notin \mathcal{R}(i) \,\}
\big]_{j \in \mathcal{N}_i^{\mathrm{hop}}},
\end{equation}
where the neighbors in $\mathcal{N}_i^{\mathrm{hop}}$ are ordered. This indicator identifies the boundary directions of
intersection $i$ and provides coarse spatial information useful for regional
coordination.

\medskip
Here the TLE determines which approaches are boundary approaches for computing
$f_{\mathcal{R}(i)}$ and identifies which directional neighbors of
intersection $i$ lie outside $\mathcal{R}(i)$ when forming
$\psi_i^{s}$.

\noindent\text{RegionWide Regional Reward $r_i^{\mathrm{regional}}$:}
In the RegionWide model, the regional reward is constructed from whole-region congestion and flow indicators aligned with the regional state design. We first form a region-averaged queue:
\begin{equation}
  \bar{q}_{\mathcal{R}(i)}^{\mathrm{mean}}(n)
  \;=\;
  \frac{1}{\lvert \mathcal{R}(i) \rvert}
  \sum_{j \in \mathcal{R}(i)} \bar{q}_j(n).
\end{equation}
Next, we reuse the regional spillback ratio $\chi_{\mathcal{R}(i)}$ and the boundary-flow indicator $f_{\mathcal{R}(i)}$ introduced in the RegionWide state definition, and introduce a lost-time estimate $\sigma_{\mathcal{R}(i)}$ defined as the fraction of intersections in $\mathcal{R}(i)$ that are currently in a yellow phase. The scalar regional reward associated with region $\mathcal{R}(i)$ at decision time $t_i^{(n)}$ as:
\begin{equation}
\begin{aligned}
  r_i^{\mathrm{regional}}
  \;=\;& -\,\bar{q}_{\mathcal{R}(i)}^{\mathrm{mean}}(n) \\
       & -\,\lambda_{\mathrm{spill}}\,\chi_{\mathcal{R}(i)}(n) \\
       & -\,\lambda_{\mathrm{switch}}\,\sigma_{\mathcal{R}(i)}(n) \\
       & +\,\lambda_{\mathrm{out}}\,f_{\mathcal{R}(i)}(n),
\end{aligned}
\label{eq:regionwide-regional-reward}
\end{equation}

\noindent which combines with $r_i^{\mathrm{local}}$ to finalize the composite reward attributed to intersection $i$ at decision step $n$.

To stabilize learning, the spillback, switching, and outflow terms are smoothed over time using an exponential moving average, while the mean-queue term is kept unsmoothed to preserve its instantaneous sensitivity to congestion.

\medskip
The regional coefficients 
$\lambda_{\mathrm{spill}}$, $\lambda_{\mathrm{switch}}$, and $\lambda_{\mathrm{out}}$
are tuned using Simultaneous Perturbation Stochastic Approximation (SPSA), a gradient-free stochastic optimization method that estimates the gradient of a noisy performance objective by randomly perturbing all parameters at once and forming a two-sided finite-difference estimate~\citep{Spall1992}. 
In our setting, the objective is defined as the long-run performance of the learned policies under stochastic traffic demand and simulation noise. We first train the regional agents in an initial, long training block, and then proceed through a sequence of alternating short training and SPSA blocks. In each SPSA block, the learning procedure is completely frozen and  for every region we generate two perturbed versions of its current coefficients in opposite directions and run paired probe episodes. The difference between the resulting regional rewards under these two oppositely perturbed coefficient settings provides a noisy gradient signal, which is then used to update the underlying coefficients in a direction that improves the average regional reward per episode. This interleaved schedule gradually adapts the regional reward weights while the policies themselves continue to train. According to the experiments, the most effective balance between local and regional terms was obtained with $\beta_{\mathcal{R}(i)} = 0.5$ for all regions.

\subsection{OneHop Model}
\label{sec:onehop}

\begin{figure}[ht]
    \centering
    \includegraphics[width=1.0\linewidth]{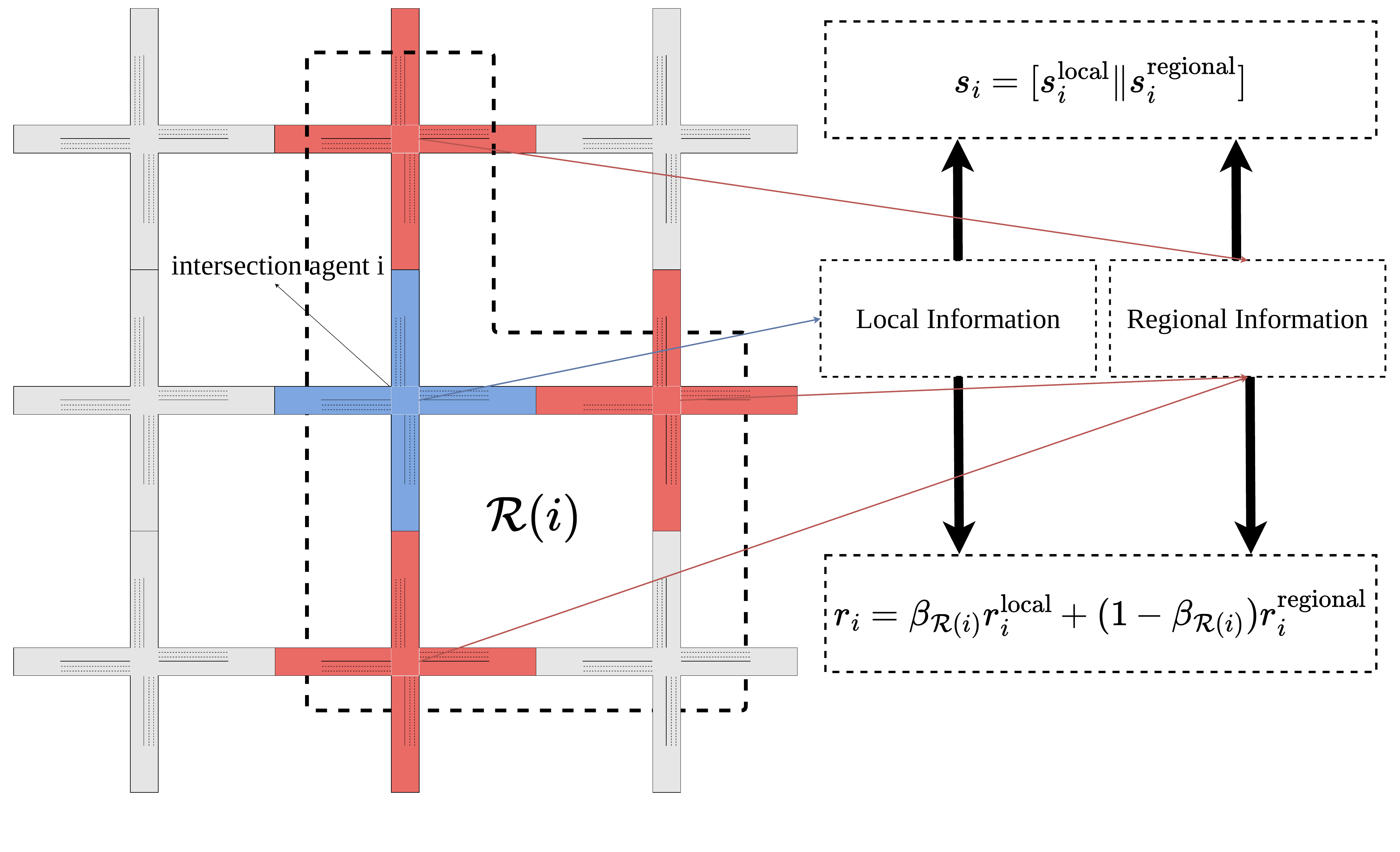}
    \caption{OneHop model receptive field.}
    \label{fig:onehop-rf}
\end{figure}

\hspace*{1em}OneHop uses bounded neighborhood receptive field to encode regional information. This scheme is presented in Fig. \ref{fig:onehop-rf}.

\noindent\text{OneHop Regional State $s_i^{\mathrm{regional}}$:}
In this model, each intersection agent $i$ exposes to its regional agent a short-range description of traffic conditions at and immediately around itself. Rather than aggregating the entire region, the OneHop design focuses exclusively on $i$ and its one-hop neighbors within the same region, enabling the regional agent to coordinate interactions driven by local congestion propagation. Consistent with the structured decomposition of $s_i^{\mathrm{regional}}$ in Eq.~\eqref{eq:regional-state}
we instantiate the regional features as follows.

\medskip
\noindent\emph{$\psi_i^{p}$:}
For each one-hop neighbor $j \in \mathcal{N}_i^{\mathrm{hop}}$ that lies in the same region, we include the neighbor’s current signal phase encoded by the one-hot vector $\text{phase}_j \in \{0,1\}^{\lvert\mathcal{A}^G\rvert}$, together with a scalar $\Delta t_j$ giving the remaining time until $j$’s next decision step.

\begin{equation}
\psi_i^{p}
=
\big[
\;\{\,\text{phase}_j,\; \Delta t_j\,\}_{j \in \mathcal{N}_i^{\mathrm{hop}}}
\big].
\end{equation}

\medskip
\noindent\emph{$\psi_i^{t}$:}
First, we define vertical and horizontal approaching flows to $i$ as
\begin{equation}
N_i^{v} = p_i^{v} - q_i^{v}\,
\qquad
N_i^{h}  =p_i^{h} - q_i^{h}.
\end{equation}

\noindent where $p_i$ and $q_i$ respectively denote the number of vehicles present and the number of vehicles halted on the immediate incoming approaches to intersection $i$, measuring how many vehicles are currently advancing toward $i$ from its neighbors. Additionally, for each neighbor $j$, we further mirror the throughput structure used in the local state by incorporating the neighbor’s halted-vehicle aggregates normalized according to $j$’s topology. 
\begin{equation}
\psi_i^{t}
\;=\;
\big[
N_i^{v},
N_i^{h},
q_j^{v},
q_j^{h},
q_j^{vl},
q_j^{hl}
\big]
\end{equation}
\medskip

\noindent\emph{$\psi_i^{s}$:}
For each one-hop neighbor we consider the structural property using the same one-hot topology encoding defined for the local spatial feature in Eq.~\eqref{eq:type-onehot}.

\medskip
Collecting these elements over all neighbors in $\mathcal{N}_i^{\mathrm{hop}}$, we obtain a fixed-layout representation even when the number of neighbors differs. Any neighbor that does not exist or lies outside the region partition is handled by the TLE, which inserts an all-zero placeholder for that neighbor’s phase, throughput, and spatial blocks.

\noindent\text{OneHop Regional Reward $r_i^{\mathrm{regional}}$:}
In the OneHop model, the regional reward complements the local regional term by evaluating congestion in the immediate neighborhood surrounding intersection $i$. Consistent with the short-range regional state design, we define the regional reward as the negative aggregate queue of all one-hop neighbors of $i$ that lie within the same region. Following the same notations introduced in~\S\ref{sec:local-reward-exact}  the regional reward attributed to $i$ at decision step $t_i^{(n)}$ is:

\begin{equation}
  r_i^{\mathrm{regional}}
  \;=\;
  - \sum_{j \in \mathcal{N}_i^{\mathrm{hop}}} \bar{q}_j(n).
\end{equation}

\noindent This formulation encourages intersection $i$ to select actions that support local–neighbor coordination, relieving spillback and stabilizing queue propagation across the immediate vicinity. The composite coefficients $(\beta_{\mathcal{R}(i)},1-\beta_{\mathcal{R}(i)})$ balancing the local and regional components are determined empirically; in all experiments, the best-performing configuration was found to be $\beta_{\mathcal{R}(i)} = 0.7$ for all regions.

\section{Experiments}
\hspace*{1em}This section details simulation setup, training details, experimental design and extensive comparison addressing the effect of SEMI-CTDE model design and its ablations. We then report quantitative results and discuss their implications.

\subsection{General Settings}
\label{sec:simulation-settings}
\hspace*{1em}Here, we describe the common simulation and training configuration that underpins all experiments.

\subsubsection{Traffic Network setting}
\label{sec:traffic-network-setting}
In this paper, we have used
the well-known open source Simulation of Urban MObility (SUMO)
simulator \citep{sumo}, to conduct our experiments.

All models are trained and evaluated on a single 5×5 grid network, including 21 signalized intersections. All approaches have uniform length (470\,m), and each approach provides three lanes. Lane-level turning permissions and right-of-way rules follow the specification in \S\ref{sec:region-based-marl-for-tsc}. The network comprises 9 cross-intersections and 12 T-intersections. A single vehicle class is used, with SUMO’s default speedFactor of 1.0 for all vehicles. Additionally, $g_l$ and $g_s$ are set to 15s and 5s, respectively.

\subsubsection{Traffic Flows}
\label{sec:traffic-flows}

The distribution of vehicle arrival times determines the traffic demand injected into the network. During training, we use a uniform arrival distribution to stabilize learning, since real-world flows often exhibit peaks that corrupt convergence. For testing, we evaluate Weibull and Gaussian arrival distributions. The simulation horizon is fixed at 18{,}000\,seconds. Table~\ref{tab:traffic-flows} summarizes the configuration of all flows used in this work.

\begin{table}[t]
    \centering
    \caption{Traffic flow configurations.}
    \label{tab:traffic-flows}
    \resizebox{\linewidth}{!}{
    \begin{tabular}{@{}lccc@{}}
        \toprule
        \textbf{Distribution} & \textbf{Total vehicles (veh)} & \textbf{Avg. rate (veh/s)} & \textbf{Name} \\
        \midrule
        Uniform       & 39{,}600 & 2.2 & U1 \\
        Gaussian      & 10{,}800 & 0.60 & G1 \\
        Gaussian      & 13{,}500 & 0.75 & G2 \\
        Weibull    & 14{,}400 & 0.60 & W1 \\
        Weibull   & 10{,}800 & 0.70 & W2 \\
        Weibull   & 18{,}000 & 1.0 & W3 \\
        Weibull   & 19{,}800 & 1.1 & W4 \\

        \bottomrule
    \end{tabular}
    }
\end{table}

\subsubsection{Training setting}
\label{sec:training-setting}

The training settings are identical across all models. All of them employ Double DQNs with a three-layer MLP backbone (hidden sizes 512–256–128) between the input and output layers. Experiences are stored in a fixed-capacity replay memory with first-in–first-out replacement: once the buffer is full, newly collected transitions overwrite the oldest ones, so each experience remains available only for a limited number of episodes. Minibatches are sampled uniformly from the buffer, and exploration follows an $\varepsilon$-greedy policy with linear decay. Training details and hyperparameters are listed in Table~\ref{tab:training-hyperparams}. 

\begin{table}[t]
    \centering
    \caption{Training details and hyperparameters.}
    \label{tab:training-hyperparams}
    \begin{tabular}{@{}p{1.2cm}>{\raggedright\arraybackslash}p{4.2cm}p{2.0cm}@{}}

        \toprule
        \textbf{Variable} & \textbf{Description} & \textbf{Value} \\
        \midrule
        $\varepsilon_{\min}$ & Minimum exploration rate & $0.01$ \\
        $\varepsilon_{\mathrm{decay}}$ & Exploration decay & $0.99$ \\
        $\eta$ & Learning rate & $2.5 \times10^{-4}$ \\
        $\gamma$ & Discount factor & $0.95$ \\
        $|B|$ & Minibatch size & $64$ \\
        $C_{\mathrm{policy}}$ & Policy network update counter  & $20$ \\
        $C_{\mathrm{target}}$ & Target network update counter & $2000$ \\
        $|D|$ & Replay memory capacity & $5 \times 10^{4}$ \\
        $\mathcal{O}$ & Optimizer & Adam \\
        $\mathcal{L}$ & Loss function & MSE \\
        \bottomrule
    \end{tabular}
\end{table}

\subsubsection{Performance Evaluation Indicators}
\label{sec:performance-evaluation-indicators}
We assess model performance using four network-wide metrics computed over and during the full simulation horizon.

Average Waiting Time (AWT): Waiting time for vehicle $v$ is the total time under a near-standstill threshold $v_{\mathrm{th}}=0.1\,\mathrm{m/s}$. AWT is the mean waiting time over all vehicles in the simulation.

Average Travel Time (ATT):
Travel time for vehicle $v$ is the elapsed time between its network entry and exit. ATT is the mean travel time over all vehicles in the simulation.

Average Queue Length (AQL): The network queue length is defined as the average number of vehicles waiting per lane across the network at time $t$. AQL is the mean queue length over the simulation horizon.

\subsection{Comparative Settings}
\label{sec:comprative-experiments}
\hspace*{1em}In order to thoroughly verify the effectiveness of models designed under SEMI-CTDE, We compare them  with other TSC algorithms as below.

\subsubsection{Rule-based control methods}
\label{sec:performance-evaluation-indicators}

Actuated (ACT) : As a conventional traffic-responsive rule-based baseline, we employ the built-in actuated traffic signal controller provided by SUMO. This controller implements a detector-based, gap-out strategy: each phase is subject to a minimum green time, may be extended while upstream detectors register continuous demand, and is terminated once a critical gap elapses or the maximum green time is reached.

\subsubsection{Fully Decentralized (FD)}
\label{sec:fully-decentralized}
This baseline follows a purely decentralized learning paradigm with no region definition, communication, or parameter sharing. Each intersection $i$ is controlled by an independent DDQN that optimizes its own signal policy using only locally observable information.

To isolate the value of SEMI-CTDE, we reuse the local components of our implementations of SEMI-CTDE, namely OneHop and RegionWide and remove all regional terms. Concretely,
\begin{equation}
s_i \;=\; s_i^{\mathrm{local}}, 
\qquad
r_i \;=\; r_i^{\mathrm{local}}.
\end{equation}

This fully decentralized baseline represents a strong, commonly used reference in multi-agent TSC: it preserves sample efficiency and stability characteristics of DDQN while foregoing any explicit coordination. Performance differences relative to SEMI-CTDE models therefore quantify the incremental benefit of region definition and parameter sharing.

\subsubsection{Partially SEMI-CTDE}
\label{sec:partially-semi-ctde}
This baseline is designed in order to address the effectiveness of composite state and reward definition and particularly regional components. This model implements all the guidelines of SEMI-CTDE including parameter sharing except the regional counterparts in state and reward formulation.
Intersection agents within each region $R_k$ share a single DDQN policy, but both the state and reward fed to the regional agent $u_k$ are strictly local and use the same definition of state and reward local components employed in our implementations of SEMI-CTDE, OneHop and RegionWide models:
\begin{equation}
s_i \;=\; s_i^{\mathrm{local}}, 
\qquad
r_i \;=\; r_i^{\mathrm{local}},
\qquad
\forall i \in R_k,
\end{equation}
which is equivalent to setting regional weights to zero. Network partitioning follows the same region definitions used throughout this work.

\subsection{Results}
\label{sec:results}
\begin{figure*}[t]
  \centering
  \begin{subfigure}[t]{0.9\textwidth}
    \centering
    \includegraphics[width=0.75\linewidth]{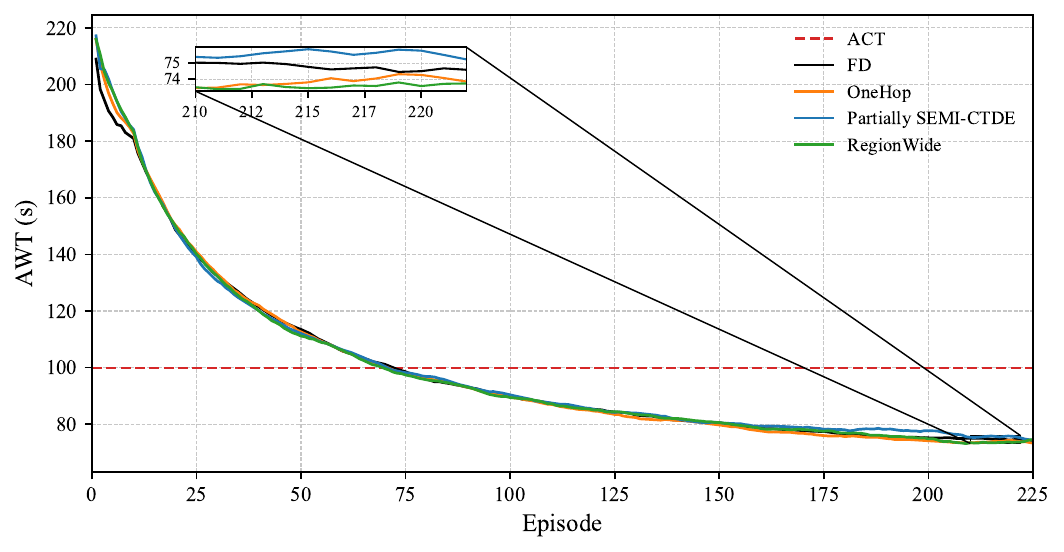}
    \caption{Average Waiting Time (AWT).}
    \label{fig:train-awt}
  \end{subfigure}

  \vspace{0.5cm}

  \begin{subfigure}[t]{0.9\textwidth}
    \centering
    \includegraphics[width=0.75\linewidth]{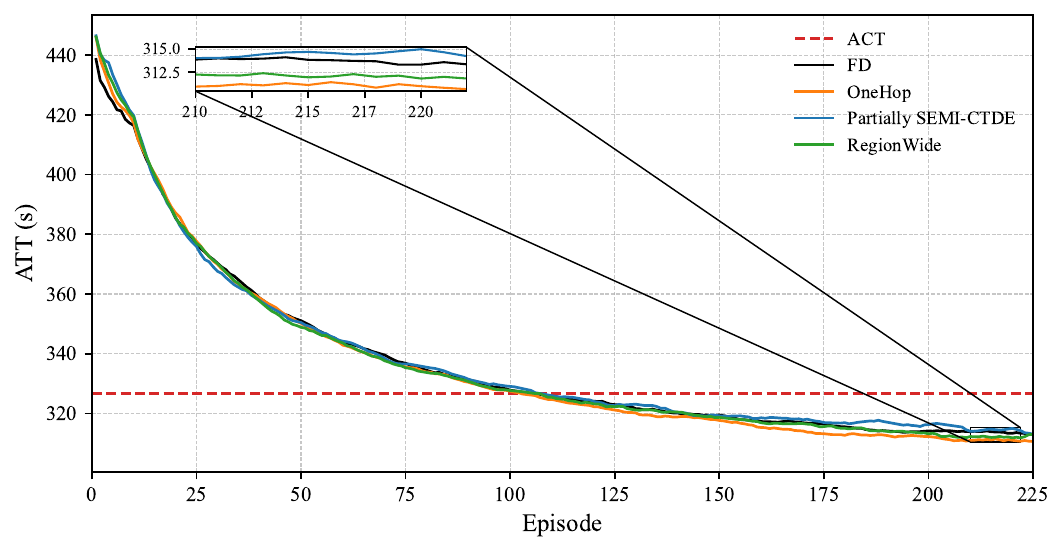}
    \caption{Average Travel Time (ATT).}
    \label{fig:train-att}
  \end{subfigure}

  \caption{10-episode moving averages of network-wide AWT and ATT over 225 episodes, with a zoom on Episodes 210–222.}
  \label{fig:training-curves}
\end{figure*}

\textbf{Training behavior:}
Fig.~\ref{fig:training-curves} tracks the evolution of network-wide AWT and ATT during training on the uniform-demand setting (U1). All learning-based policies improve steadily. Curves descend and then stabilize as exploration decays. For ATT, Partially~SEMI-CTDE exhibits an early advantage between Episodes~25–40, which is likely due to its smaller, simpler state and reward definitions. Yet from roughly Episode~100 onward, OneHop consistently dominates and maintains a clear margin through the end of training. RegionWide is the second-best performer and, after convergence, opens a clear gap over both FD and Partially~SEMI-CTDE. For AWT, Partially~SEMI-CTDE again leads in the early stages, but from around Episode~120 OneHop becomes dominant. After Episode~200, RegionWide surpasses OneHop, and both maintain a pronounced margin over the other two baselines. A consistent observation is that FD outperforms Partially~SEMI-CTDE on both metrics: because they share identical local state and reward while FD trains a separate network per intersection, Partially~SEMI-CTDE’s parameter sharing reduces capacity without adding regional terms, indicating that parameter sharing alone does not guarantee gains and that state/reward design is equally critical. As is visible in both metrics, the SEMI-CTDE models (RegionWide, OneHop) outperform the other models, and this confirms that regional state/reward alongside coordination via parameter sharing improves learning.

\textbf{Test-time performance:}
To better analyze and compare the results, we group the tests based on the intensity of the arrival distributions. Each model is tested 10 times with different random seeds for each flow. We report the mean and standard deviation of the model performance in AQL and AWT(Tables~\ref{tab:aql-light}–\ref{tab:awt-heavy}).

\begin{table*}[t]
  \centering

  \begin{minipage}{0.48\linewidth}
  \Large
  \caption{AQL (m) - Light flows (10 runs)}
    \centering
    \label{tab:aql-light}
    {\renewcommand{\arraystretch}{1.6}%
    \resizebox{\linewidth}{!}{%
    \begin{tabular}{@{}l S S S S S@{}}
      \toprule
      {Flow} & {RegionWide} & {OneHop} & {Partially SEMI-CTDE} & {FD} & {Actuated} \\
      \midrule
      W1\       & \num{7.67 \pm 0.05}  & \best{\num{6.90 \pm 0.04}} & \num{7.19 \pm 0.08} & \num{7.60 \pm 0.07} & \num{7.52 \pm 0.00} \\
      W2\ &\num{8.01 \pm 0.09} & \best{\num{7.21 \pm 0.02}}  & \num{7.54 \pm 0.04} & \num{7.91 \pm 0.05}  & \num{8.20 \pm 0.00} \\
      G1\ & \num{8.46 \pm 0.12}        & \best{\num{7.86 \pm 0.10}}  & \num{8.26\pm 0.09} & \num{8.52 \pm 0.18} & \num{8.92 \pm 0.00} \\
      \bottomrule
    \end{tabular}}}
  \end{minipage}
  \hfill
  \begin{minipage}{0.48\linewidth}
  \Large
    \caption{AWT(s)- Light flows (10 runs)}
    \centering
    \label{tab:awt-light}
    {\renewcommand{\arraystretch}{1.6}%
    \resizebox{\linewidth}{!}{%
    \begin{tabular}{@{}l S S S S S@{}}
      \toprule
      {Flow} & {RegionWide} & {OneHop} & {Partially SEMI-CTDE} & {FD} & {Actuated} \\
      \midrule
      W1  & \num{38.35 \pm 0.50} & \best{\num{28.27 \pm 0.22}} & \num{32.35 \pm 0.49} & \num{35.73 \pm 0.41} & \num{34.31 \pm 0.00} \\
      W2  & \num{40.36 \pm 0.74} & \best{\num{30.33 \pm 0.20}} & \num{34.28 \pm 0.33} & \num{37.12 \pm 0.28} & \num{38.95 \pm 0.00} \\
      G1\ & \num{47.38 \pm 0.99}  & \best{\num{38.00 \pm 0.65}}  & \num{43.16 \pm 0.70} & \num{45.17 \pm 0.96} & \num{50.11 \pm 0.00} \\
      \bottomrule
    \end{tabular}}}
  \end{minipage}
\end{table*}

\begin{table*}[t]
  \centering

  \begin{minipage}{0.48\linewidth}
  \Large
   \caption{AQL (m) - Heavy flows (10 runs)}
    \centering
    \label{tab:aql-heavy}
    {\renewcommand{\arraystretch}{1.6}%
    \resizebox{\linewidth}{!}{%
    \begin{tabular}{@{}l S S S S S@{}}
      \toprule
      {Flow} & {RegionWide} & {OneHop} & {Partially SEMI-CTDE} & {FD} & {Actuated} \\
      \midrule
      W3\          & \best{\num{11.54 \pm 0.16}}  & \num{13.07 \pm 0.92}  & \num{13.59 \pm 0.23} & \num{12.34 \pm 0.14} & \num{12.96 \pm 0.00} \\
      W4\   & \best{\num{13.26 \pm 0.19}}  & \num{17.60 \pm 1.49} & \num{17.21 \pm 0.48} & \num{117.16 \pm 111.39} & \num{15.07 \pm 0.00} \\
      G2\ & \best{\num{10.67 \pm 0.17}} & \num{11.68 \pm 0.44} & \num{12.15 \pm 0.14} & \num{11.39 \pm 0.29} & \num{11.53 \pm 0.00} \\
      \bottomrule
    \end{tabular}}}
  \end{minipage}
  \hfill
  \begin{minipage}{0.48\linewidth}
  \Large
    \caption{AWT (s) - Heavy flows (10 runs)}
    \centering
    \label{tab:awt-heavy}
    {\renewcommand{\arraystretch}{1.6}%
    \resizebox{\linewidth}{!}{%
    \begin{tabular}{@{}l S S S S S@{}}
      \toprule
      {Flow} & {RegionWide} & {OneHop} & {Partially SEMI-CTDE} & {FD} & {Actuated} \\
      \midrule
      W3\     & \best{\num{65.38 \pm 1.22}}  & \num{67.61 \pm 3.19} & \num{72.91 \pm 0.98} & \num{66.85 \pm 0.92} & \num{76.05 \pm 0.00} \\
      W4\   & \best{\num{75.99 \pm 1.51}}  & \num{95.48 \pm 8.03}& \num{94.11 \pm 3.54} & \num{400.42 \pm 213.56} & \num{91.33 \pm 0.00} \\
      G2\ & \best{\num{65.92 \pm 0.94}} & \num{67.65 \pm 2.96} & \num{72.34 \pm 1.02} & \num{67.93 \pm 1.97} & \num{74.80 \pm 0.00} \\
      \bottomrule
    \end{tabular}}}
  \end{minipage}

\end{table*}

Under light-demand arrival distributions (Tables~\ref{tab:aql-light}–\ref{tab:awt-light}), OneHop consistently achieves the lowest average queues and waiting times among all models. This is expected because, in these distributions, congestion is mild, queues remain short, and interactions are dominated by the immediate upstream and downstream neighbors of each intersection rather than by far-away parts of the region. OneHop’s bounded neighborhood design exposes exactly this local coupling: the regional terms summarize only the immediate influential vicinity, so the regional agent receives just enough information to coordinate adjacent intersection agents efficiently without being distracted by weakly relevant fluctuations elsewhere. In contrast, RegionWide aggregates state and reward information over the entire region, mixing lightly loaded and moderately loaded intersections into a single coarse summary. When flows are light, these whole-region aggregates carry little signal about the few locations where temporary queues actually form, and they can even be misleading by smoothing out localized imbalances that the regional agent should react to. As a result, the learning signal for RegionWide becomes noisy and poorly aligned with the truly critical decisions at each intersection agent, leading to systematically worse AQL and AWT despite using the same network capacity as OneHop. Finally, the light-flow results also show that Partially SEMI-CTDE consistently outperforms the FD baseline. Both rely purely on local information, but Partially SEMI-CTDE shares a single DDQN per region, so each update is trained on experience collected from multiple similar intersections. This parameter sharing increases sample efficiency, stabilizes learning, and yields smoother, more coherent signal policies across the region, whereas independent learners in the FD setup must each generalize from a much smaller, noisier local experience stream, which hurts their ability to keep queues and delays low even in these relatively easy traffic conditions.

Under heavy-demand arrival distributions (Tables~\ref{tab:aql-heavy}–\ref{tab:awt-heavy}), RegionWide clearly dominates all other controllers, indicating that whole-region aggregation becomes crucial once congestion intensifies and queues begin to interact across wide spatial scales. In these distributions, bottlenecks no longer arise in isolation: spillback from one intersection can quickly propagate through corridors and across the region, and RegionWide’s design, which pools information over all intersections in the region, gives each regional agent a faithful picture of these global imbalances. This richer regional information allows the regional agent to coordinate actions in a way that proactively protects critical corridors and clears queues before they cascade, which is reflected in RegionWide’s consistently superior queues and delays across all heavy flows. The most striking case is W4, the heaviest demand pattern in our tests, where RegionWide not only maintains stable performance but opens a very large margin over all other learning-based methods, demonstrating strong generalization under extreme arrivals. In contrast, OneHop performs poorly in these distributions: its bounded neighborhood regional terms, which were ideal under light flows, become insufficient once congestion and spillback span multiple intersections, so the regional agent reacts shortsightedly to local vicinity rather than the region-scale structure of queues. Nevertheless, OneHop still outperforms Partially SEMI-CTDE, which lacks regional information altogether and therefore cannot exploit any cross intersection coordination, underscoring that even limited regional state can be helpful when combined with parameter sharing. FD exhibits an interesting trade-off: in most heavy flows where it runs, its larger parameter capacity (a separate DDQN per intersection) yields slightly better performance than OneHop, suggesting that, in the absence of strong regional signals, sheer model capacity can partially compensate. However, in W4 the FD model fails entirely to mitigate traffic congestion, whereas other algorithms remain stable and effective, reinforcing the importance of region-based design and shared parameters for robustness under extreme congestion. Finally, the actuated baseline is consistently outperformed by all learning-based models. In heavy flows, the gap in AWT and AQL becomes substantial, but even in light flows the learning-based policies retain a smaller yet persistent advantage, indicating that data-driven coordination yields meaningful gains over traditional detector-based control across the whole range of arrival distributions.

To complement the aggregate, episode-level statistics reported in Tables~\ref{tab:aql-light}–\ref{tab:awt-heavy}, we next examine how network performance evolves over simulation time within representative frozen test episodes drawn from the very same runs used to compute these tables. Instead of summarizing each run by a single scalar, these plots track the trajectories of queue length and waiting time of the network. These temporal perspectives, along with waiting time histograms (Figs.~\ref{fig:6-light}–\ref{fig:6-heavy}) reveals how each model responds to demand peaks, how it recovers from it and let us compare their performances during different arrival rates, thereby providing a more fine-grained view of the mechanisms underlying the performance gaps observed in the tables.

\newcommand{\sixpdfwidth}{\linewidth} 
\begin{figure*}[t]
  \centering
  
  \begin{subfigure}[t]{0.49\textwidth}
    \centering
    \begin{subfigure}[t]{\textwidth}
      \centering
      \includegraphics[width=\sixpdfwidth]{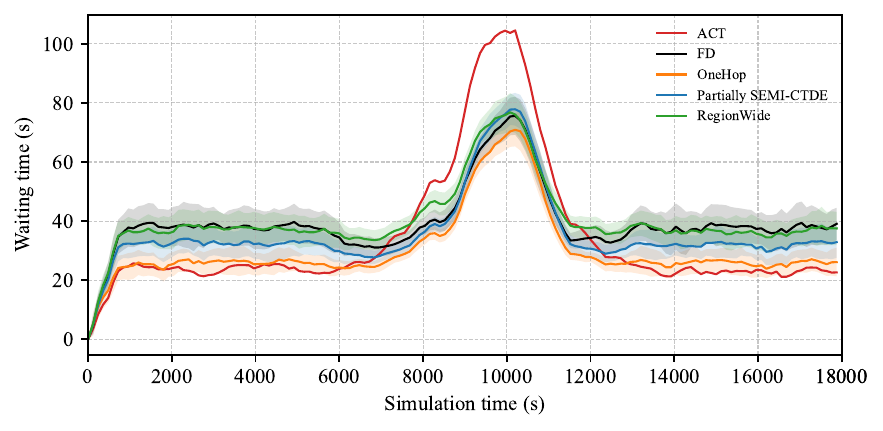}
      \caption{}
      \label{fig:g1-awt}
    \end{subfigure}
    \\[2pt]
    \begin{subfigure}[t]{\textwidth}
      \centering
      \includegraphics[width=\sixpdfwidth]{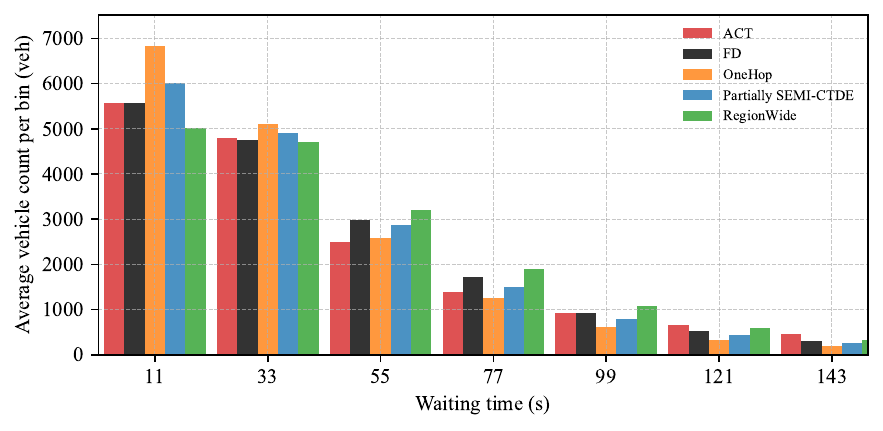}
      \caption{}
      \label{fig:g1-hist}
    \end{subfigure}
    \\[2pt]
    \begin{subfigure}[t]{\textwidth}
      \centering
      \includegraphics[width=\sixpdfwidth]{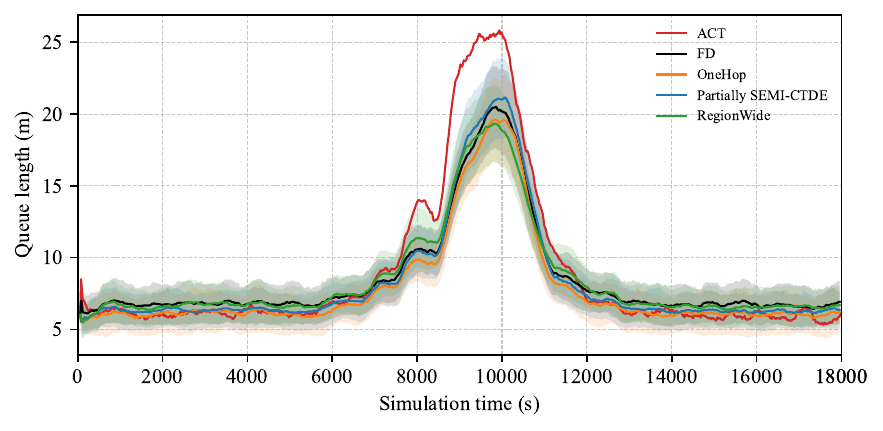}
      \caption{}
      \label{fig:g1-aql}
    \vspace{10pt}
    \end{subfigure}
    \centering\textbf{Flow G1}
  \end{subfigure}\hfill
  %
  \begin{subfigure}[t]{0.49\textwidth}
    \centering
    \begin{subfigure}[t]{\textwidth}
      \centering
      \includegraphics[width=\sixpdfwidth]{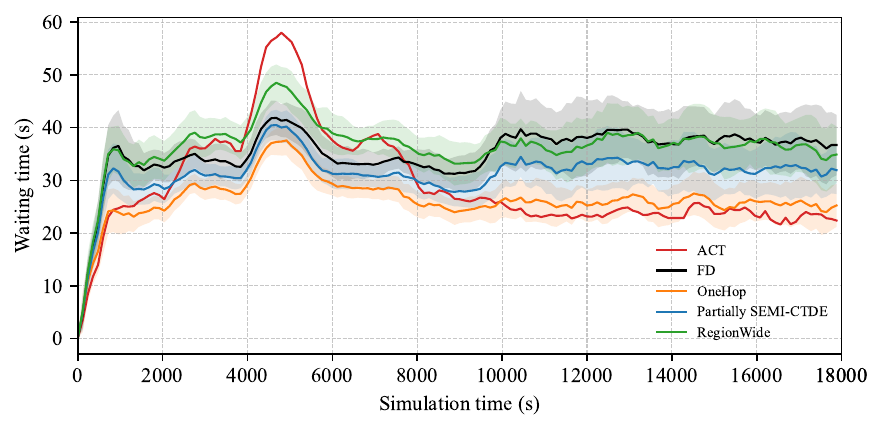}
      \caption{}
      \label{fig:w1-awt}
    \end{subfigure}
    \\[2pt]
    \begin{subfigure}[t]{\textwidth}
      \centering
      \includegraphics[width=\sixpdfwidth]{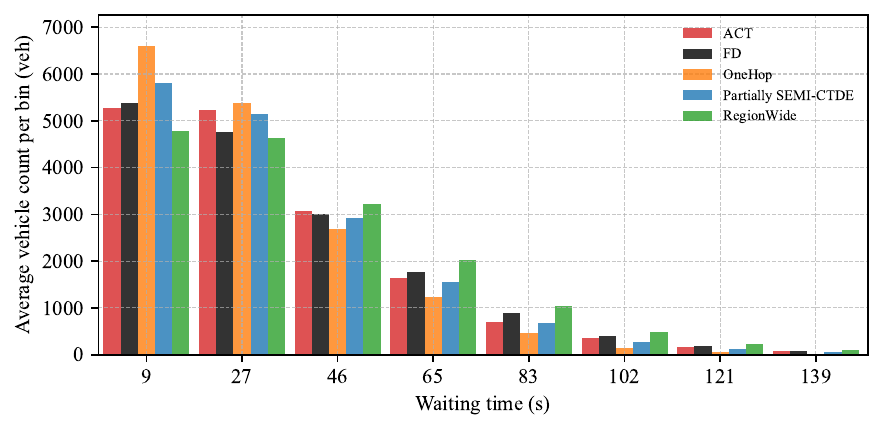}
      \caption{}
      \label{fig:w1-hist}
    \end{subfigure}
    \\[2pt]
    \begin{subfigure}[t]{\textwidth}
      \centering
      \includegraphics[width=\sixpdfwidth]{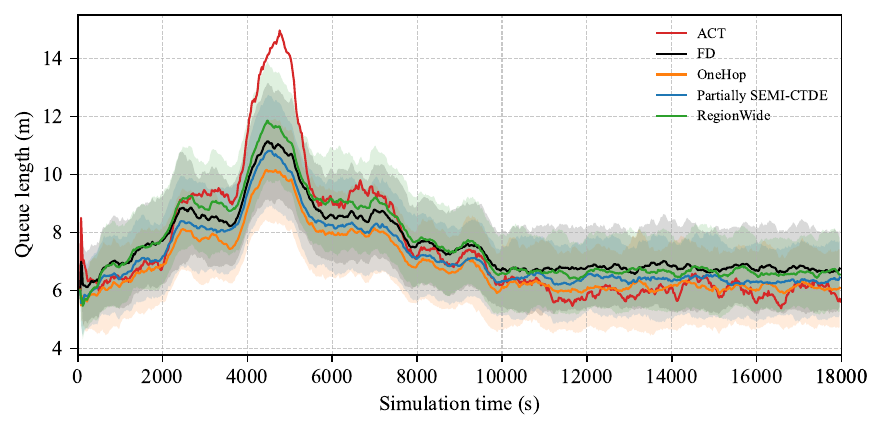}
      \caption{}
      \label{fig:w1-aql}
      \vspace{10pt}
    \end{subfigure}
    \vspace{10pt}
    \centering\textbf{Flow W1}
  \end{subfigure}
  \caption{Intra-episode performance under light-demand arrivals. Figures (a–c) correspond to Flow G1 and figures (d–f) to Flow W1. (a,d) show the moving average waiting time; (b,e) show waiting time histograms. Bar height is the average vehicle count per bin; (c,f) show the moving average queue length. All figures are based on the same set of 10 runs in Tables~\ref{tab:aql-light}–\ref{tab:awt-light}.}
  \label{fig:6-light}
\end{figure*}

\begin{figure*}[t]
  \centering
  
  \begin{subfigure}[t]{0.49\textwidth}
    \centering
    \begin{subfigure}[t]{\textwidth}
      \centering
      \includegraphics[width=\sixpdfwidth]{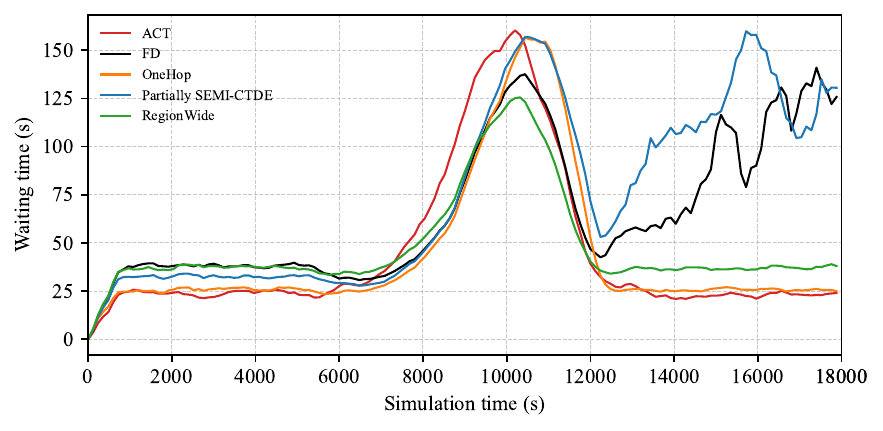}
      \caption{}
      \label{fig:g2-awt}
    \end{subfigure}
    \\[2pt]
    \begin{subfigure}[t]{\textwidth}
      \centering
      \includegraphics[width=\sixpdfwidth]{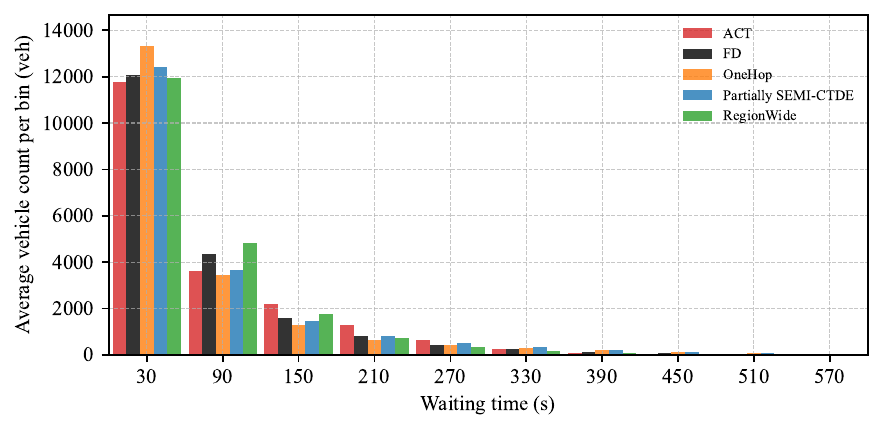}
      \caption{}
      \label{fig:g2-hist}
    \end{subfigure}
    \\[2pt]
    \begin{subfigure}[t]{\textwidth}
      \centering
      \includegraphics[width=\sixpdfwidth]{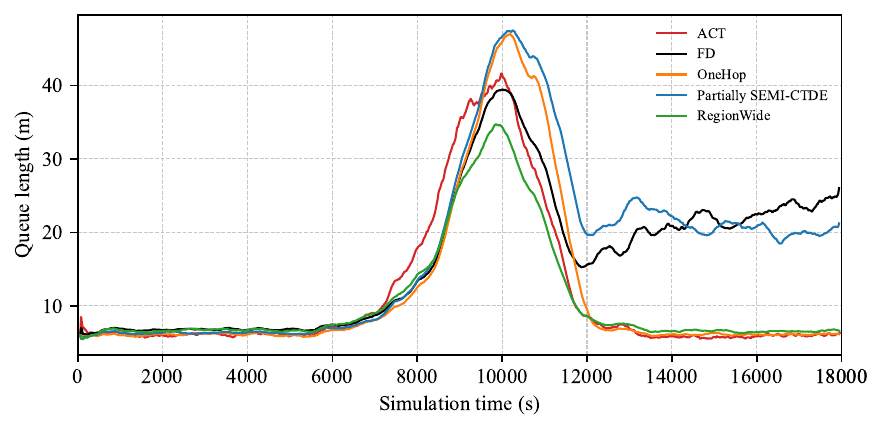}
      \caption{}
      \label{fig:g2-aql}
    \vspace{10pt}
    \end{subfigure}
    \centering\textbf{Flow G2}
  \end{subfigure}\hfill
  %
  \begin{subfigure}[t]{0.49\textwidth}
    \centering
    \begin{subfigure}[t]{\textwidth}
      \centering
      \includegraphics[width=\sixpdfwidth]{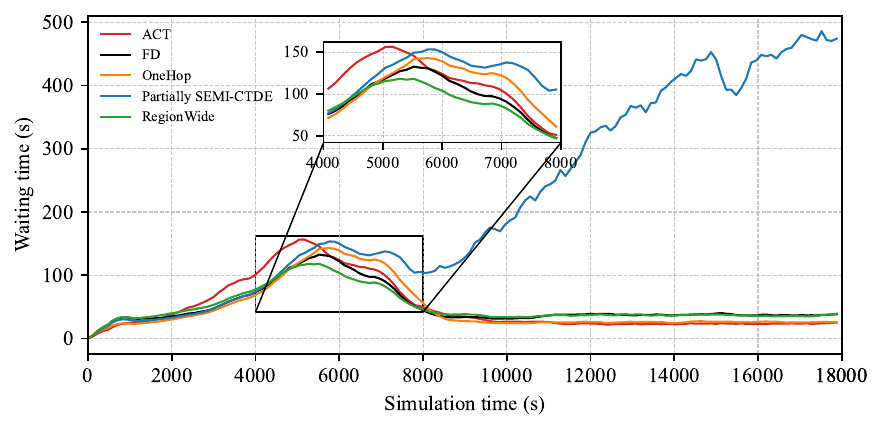}
      \caption{}
      \label{fig:w3-awt}
    \end{subfigure}
    \\[2pt]
    \begin{subfigure}[t]{\textwidth}
      \centering
      \includegraphics[width=\sixpdfwidth]{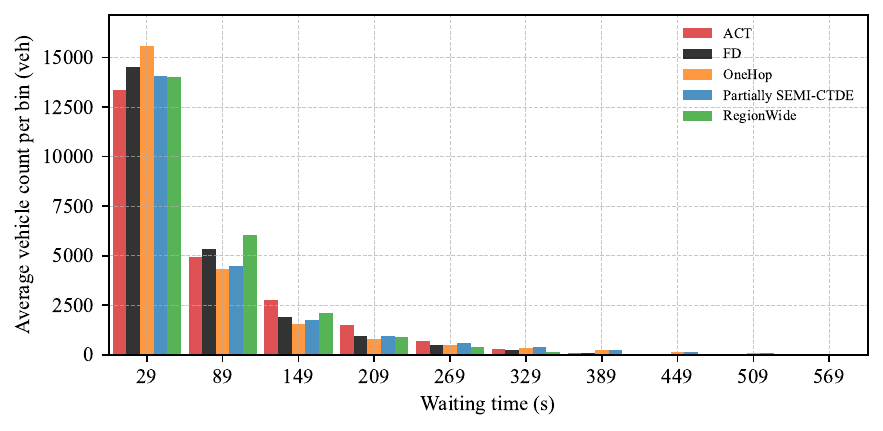}
      \caption{}
      \label{fig:w3-hist}
    \end{subfigure}
    \\[2pt]
    \begin{subfigure}[t]{\textwidth}
      \centering
      \includegraphics[width=\sixpdfwidth]{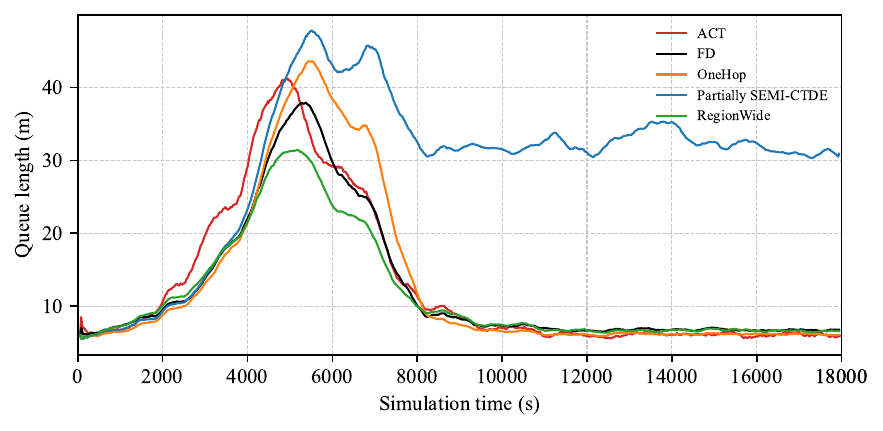}
      \caption{}
      \label{fig:w3-aql}
      \vspace{10pt}
    \end{subfigure}
    \vspace{10pt}
    \centering\textbf{Flow W3}
  \end{subfigure}
  \caption{Intra-episode performance under heavy-demand arrivals. Figures (a–c) correspond to Flow G2 and figures (d–f) to Flow W3. (a,d) show the moving average waiting time; (b,e) show waiting time histograms. Bar height is the average vehicle count per bin; (c,f) show the moving average queue length. All figures are based on the same set of 10 runs in Tables~\ref{tab:aql-heavy}–\ref{tab:awt-heavy}.}
  \label{fig:6-heavy}
\end{figure*}

For the light-demand flows (G1, W1), the temporal plots and waiting-time histograms reinforce the same hypothesis as the aggregate tables while revealing more detail about how each model behaves over a full episode. In the Gaussian flow G1, OneHop clearly dominates the other learning-based models in both queue length and waiting time traces (Figs.~\ref{fig:g1-aql}, \ref{fig:g1-awt}): it keeps queues and delays lowest not only around the peak but also in the lower-demand periods before and after, indicating that its bounded-neighborhood regional information is exactly tuned to the spatial scale of interactions in these light arrival distributions. The actuated model, by contrast, performs particularly poorly during the peak, with pronounced spikes in both waiting time and queue length, which explains its consistently inferior episode-level averages; however, its trajectories before and after the peak are slightly better than OneHop, suggesting that a simple rule-based strategy can still be competitive when arrival rates are very low. In G1, Partially SEMI-CTDE maintains lower queues and waiting times than both FD and RegionWide in the off-peak periods, in line with the table results and confirming that regional parameter sharing improves sample efficiency even when only local information is used. Around the peak, these three models show broadly similar behavior, all lagging behind OneHop, but the queue length curves reveal that RegionWide starts to close the gap and can even slightly surpass OneHop in the very center of the Gaussian peak, hinting that whole-region aggregation becomes more helpful once instantaneous arrival approaches heavier rates. The Weibull flow W1 (Figs.~\ref{fig:w1-aql}, \ref{fig:w1-awt}) mirrors most of these patterns with a key difference at peak: OneHop again yields the lowest queues and waiting times and actuated again performs worst, but now RegionWide is clearly inferior to both FD and Partially SEMI-CTDE at the peak. This can be explained by the fact that the peak in G1 is more intense than the peak in W1, confirming that it is only under genuinely heavier arrivals that whole-region summaries inject useful, relevant information that improves the model’s decision making. Finally, the waiting-time histograms for both G1 and W1 (Figs.~\ref{fig:g1-hist}, \ref{fig:w1-hist}) show that OneHop concentrates a much larger fraction of vehicles in the low-wait bins, with markedly fewer vehicles suffering long delays, whereas the other models exhibit heavier tails. This distributional view confirms that OneHop not only improves average performance but also delivers a better individual driving experience by reducing the likelihood of vehicles experiencing prolonged waiting times.

For the heavy-demand flows (G2, W3), the temporal plots again refine and support the aggregate picture from the tables, while clarifying when and why each model succeeds or fails. In the Gaussian flow G2 (Figs.~\ref{fig:g2-aql}, \ref{fig:g2-awt}), the waiting time curves show that before the peak, both OneHop and the actuated models maintain the lowest delays, consistent with our earlier observation that OneHop is well suited to moderate arrivals and that a simple rule-based strategy can be competitive at low demand. In pre-peak arrivals, Partially SEMI-CTDE tracks these two reasonably well and clearly outperforms both FD and RegionWide, as expected from its parameter sharing and purely local information. Around the peak, however, the ordering reverses: RegionWide becomes the clear best performer, keeping waiting time lowest, while OneHop, Partially SEMI-CTDE, and actuated all exhibit much higher delays; FD sits in between, worse than RegionWide but better than other three, likely because its larger number of learnable parameters grants it stronger representational capacity in the high-demand arrival rates. The most striking behavior occurs after the peak: Partially SEMI-CTDE and FD are unable to discharge the accumulated congestion and their waiting time and queue length curves remain high and even become worse, whereas OneHop and actuated recover quickly and achieve the best post-peak performance, with RegionWide trailing slightly behind them. This pattern suggests that the strong advantage of RegionWide in the episode-level metrics comes primarily from its ability to manage the peak itself; outside the peak it is not the best performer, but its superior control during the most congested period dominates the aggregate outcome. The queue length trajectories for G2 follow a similar pattern: RegionWide dominates at the peak, the gap to OneHop narrows after the peak, and OneHop and Partially SEMI-CTDE can even fall below actuated when regional information is absent, underscoring that in genuinely heavy demand, the presence and scale of regional information are critical. In W3 (Figs.~\ref{fig:w3-aql}, \ref{fig:w3-awt}), the qualitative picture remains the same: RegionWide again dominates at the peak, and OneHop together with actuated provide the best recovery after the peak, but here only Partially SEMI-CTDE fails to fully recover while FD manages to discharge more effectively than in G2, with all other relative comparisons preserved. The waiting-time histograms for G2 and W3 (Figs.~\ref{fig:g2-hist}, \ref{fig:w3-hist}) further clarify these effects: the lowest waiting-time bin is dominated by OneHop, with actuated contributing the smallest fraction there, which aligns with our conclusion that OneHop is particularly well suited for lower and moderate arrival rates. The second bin is dominated by RegionWide, reflecting its behavior at peak: during the heaviest intervals, RegionWide is essentially the only model that keeps network-wide delays at a moderate and manageable level rather than allowing them to explode, so many vehicles accumulate in this intermediate waiting range. In subsequent bins corresponding to larger waiting times, RegionWide exhibits the lowest frequencies among all learning-based policies, confirming that under heavy flows it not only controls peak congestion more effectively but also substantially reduces the probability of very long individual delays.

Taken together, our detailed, multi-perspective analysis—spanning aggregate episode metrics, stepwise temporal profiles, and waiting-time distributions—shows that SEMI-CTDE with carefully designed regional information delivers robust improvements across a wide range of traffic configurations. OneHop is the most effective in light and moderately loaded traffic flows, where bounded-neighborhood summaries provide just the right level of coordination, while RegionWide is the best choice under heavy congestion, where whole-region aggregation is needed to manage spillback and prevent collapse. These all confirm the value of regional parameter sharing, which highlights that region-centric semi-centralized learning is key to achieving superior network efficiency and less intense delay distributions.

\section{Conclusion}
\label{sec:conclusion}

In this work, we formulated urban traffic signal control as a region-based multi-agent reinforcement learning problem and introduced \emph{SEMI-CTDE} architecture, a semi-centralized training, decentralized execution approach tailored to this setting. Beyond this, we provided a self-contained problem formulation for region-based MARL in TSC, explicitly formalizing regions, intersection agents, and regional agents together with their associated decision processes, composite state space and reward functions designed specifically to operate in region based environments. To the best of our knowledge, this is the first work to articulate these components within a unified region-centric paradigm, which can serve as a reusable foundation for subsequent research on coordinated traffic signal control. To obtain regions, we employed a fuzzy-graph $\alpha$-cut method to extract regions whose intersections are tightly interdependent in terms of traffic features, thereby aligning the learning architecture with the dominant flow dependencies in the network. The overall architecture combines topology and location-aware encoders, admissibility-aware action mapping, asynchronous event-driven experience gathering, and regional parameter sharing via DDQN-based agents, enabling semi-centralized learning at region scope with fully decentralized execution. We instantiated this architecture with two concrete models, RegionWide and OneHop, that share all SEMI-CTDE backbones and differ only in the receptive field used for their regional components: whole-region summaries in RegionWide versus bounded one-hop neighborhoods in OneHop.

Comprehensive experiments on the 5×5 grid network under Gaussian and Weibull arrival distributions, structured around targeted comparisons, examined (i) the importance of composite local–regional state and reward design, (ii) the effect of regional parameter sharing against a fully decentralized baseline. The results closely align with the main ideas behind SEMI-CTDE: grouping tightly interdependent intersections into regions and coupling this regioning with composite state and reward formulations yields consistently superior performance compared with purely local learning and conventional rule-based control. The results further revealed a clear relation between receptive-field choice and arrival flow demand: OneHop is most effective under light and moderately demanded flows while RegionWide dominates under heavy flows. Together, these findings highlight that region-centric semi-centralized learning combined with carefully structured state and reward designs yields robust, interpretable gains over both purely local learning and traditional rule-based control.

Despite these contributions, our study has several limitations that suggest avenues for future work. First, the implemented models consider a restricted discrete action space built from four logics with short/long durations; more complex phasing paradigms, as well as continuously variable green durations, are not explored here even though SEMI-CTDE itself can naturally accommodate them. Second, our state design relies on hand-crafted spatial and temporal aggregates. While this yields interpretability, it may underutilize the expressive power of modern deep models. A natural extension is to integrate learned feature extractors such as graph neural networks to capture finer-grained spatial structure over the road network and LSTM, or other sequence models to encode temporal dynamics. These extensions are all compatible with the SEMI-CTDE backbone, replacing or augmenting the manually designed feature blocks. Exploring these richer feature spaces, together with more general action parameterizations and larger-scale or real-world networks, forms a promising direction for future research on region-based multi-agent traffic signal control under the SEMI-CTDE paradigm.

\bibliographystyle{cas-model2-names}

\bibliography{cas-refs}

\begin{thebibliography}{42}
\expandafter\ifx\csname natexlab\endcsname\relax\def\natexlab#1{#1}\fi
\providecommand{\url}[1]{\texttt{#1}}
\providecommand{\href}[2]{#2}
\providecommand{\path}[1]{#1}
\providecommand{\DOIprefix}{doi:}
\providecommand{\ArXivprefix}{arXiv:}
\providecommand{\URLprefix}{URL: }
\providecommand{\Pubmedprefix}{pmid:}
\providecommand{\doi}[1]{\href{http://dx.doi.org/#1}{\path{#1}}}
\providecommand{\Pubmed}[1]{\href{pmid:#1}{\path{#1}}}
\providecommand{\bibinfo}[2]{#2}
\ifx\xfnm\relax \def\xfnm[#1]{\unskip,\space#1}\fi
\bibitem[{Abdoos(2021)}]{9109698}
\bibinfo{author}{Abdoos, M.}, \bibinfo{year}{2021}.
\newblock \bibinfo{title}{Fuzzy graph and collective multiagent reinforcement learning for traffic signals control}.
\newblock \bibinfo{journal}{IEEE Intelligent Systems} \bibinfo{volume}{36}, \bibinfo{pages}{48--55}.
\newblock \DOIprefix\doi{10.1109/MIS.2020.3000180}.
\bibitem[{Abdoos et~al.(2011)Abdoos, Mozayani and Bazzan}]{6083114}
\bibinfo{author}{Abdoos, M.}, \bibinfo{author}{Mozayani, N.}, \bibinfo{author}{Bazzan, A.L.C.}, \bibinfo{year}{2011}.
\newblock \bibinfo{title}{Traffic light control in non-stationary environments based on multi agent q-learning}, in: \bibinfo{booktitle}{2011 14th International IEEE Conference on Intelligent Transportation Systems (ITSC)}, pp. \bibinfo{pages}{1580--1585}.
\newblock \DOIprefix\doi{10.1109/ITSC.2011.6083114}.
\bibitem[{Abdulhai et~al.(2003)Abdulhai, Pringle and Karakoulas}]{Abdulhai2003}
\bibinfo{author}{Abdulhai, B.}, \bibinfo{author}{Pringle, R.}, \bibinfo{author}{Karakoulas, G.J.}, \bibinfo{year}{2003}.
\newblock \bibinfo{title}{Reinforcement learning for true adaptive traffic signal control}.
\newblock \bibinfo{journal}{Journal of Transportation Engineering} \bibinfo{volume}{129}, \bibinfo{pages}{278–285}.
\newblock \DOIprefix\doi{10.1061/(ASCE)0733-947X(2003)129:3(278)}.
\bibitem[{Ault et~al.(2020)Ault, Hanna and Sharon}]{Ault2020}
\bibinfo{author}{Ault, J.}, \bibinfo{author}{Hanna, J.P.}, \bibinfo{author}{Sharon, G.}, \bibinfo{year}{2020}.
\newblock \bibinfo{title}{Learning an interpretable traffic signal control policy}, in: \bibinfo{booktitle}{Proceedings of the 19th International Conference on Autonomous Agents and Multiagent Systems (AAMAS 2020)}, p. \bibinfo{pages}{88–96}.
\newblock \DOIprefix\doi{10.5555/3407923.3407957}.
\bibitem[{Bao et~al.(2023)Bao, Wu, Lin, Zhong, Chen and Yin}]{Bao2023}
\bibinfo{author}{Bao, J.}, \bibinfo{author}{Wu, C.}, \bibinfo{author}{Lin, Y.}, \bibinfo{author}{Zhong, L.}, \bibinfo{author}{Chen, X.}, \bibinfo{author}{Yin, R.}, \bibinfo{year}{2023}.
\newblock \bibinfo{title}{A scalable approach to optimize traffic signal control with federated reinforcement learning}.
\newblock \bibinfo{journal}{Scientific Reports} \bibinfo{volume}{13}, \bibinfo{pages}{19184}.
\newblock \DOIprefix\doi{10.1038/s41598-023-46074-3}.
\bibitem[{Bie et~al.(2024)Bie, Ji and Ma}]{BIE2024104663}
\bibinfo{author}{Bie, Y.}, \bibinfo{author}{Ji, Y.}, \bibinfo{author}{Ma, D.}, \bibinfo{year}{2024}.
\newblock \bibinfo{title}{Multi-agent deep reinforcement learning collaborative traffic signal control method considering intersection heterogeneity}.
\newblock \bibinfo{journal}{Transportation Research Part C: Emerging Technologies} \bibinfo{volume}{164}, \bibinfo{pages}{104663}.
\newblock \URLprefix \url{https://www.sciencedirect.com/science/article/pii/S0968090X24001840}, \DOIprefix\doi{https://doi.org/10.1016/j.trc.2024.104663}.
\bibitem[{Bouktif et~al.(2023)Bouktif, Cheniki, Ouni and El{-}Sayed}]{Bouktif2023}
\bibinfo{author}{Bouktif, S.}, \bibinfo{author}{Cheniki, A.}, \bibinfo{author}{Ouni, A.}, \bibinfo{author}{El{-}Sayed, H.}, \bibinfo{year}{2023}.
\newblock \bibinfo{title}{Deep reinforcement learning for traffic signal control with consistent state and reward design approach}.
\newblock \bibinfo{journal}{Knowledge-Based Systems} \bibinfo{volume}{267}, \bibinfo{pages}{110440}.
\newblock \DOIprefix\doi{10.1016/j.knosys.2023.110440}.
\bibitem[{Cai et~al.(2025)Cai, Fang and Xu}]{CAI2025126938}
\bibinfo{author}{Cai, S.}, \bibinfo{author}{Fang, J.}, \bibinfo{author}{Xu, M.}, \bibinfo{year}{2025}.
\newblock \bibinfo{title}{Xlight: An interpretable multi-agent reinforcement learning approach for traffic signal control}.
\newblock \bibinfo{journal}{Expert Systems with Applications} \bibinfo{volume}{273}, \bibinfo{pages}{126938}.
\newblock \URLprefix \url{https://www.sciencedirect.com/science/article/pii/S0957417425005603}, \DOIprefix\doi{https://doi.org/10.1016/j.eswa.2025.126938}.
\bibitem[{Casas(2017)}]{Casas2017}
\bibinfo{author}{Casas, N.}, \bibinfo{year}{2017}.
\newblock \bibinfo{title}{Deep deterministic policy gradient for urban traffic light control}.
\newblock \bibinfo{journal}{arXiv preprint arXiv:1703.09035} .
\bibitem[{Chu et~al.(2020)Chu, Wang, Codecà and Li}]{8667868}
\bibinfo{author}{Chu, T.}, \bibinfo{author}{Wang, J.}, \bibinfo{author}{Codecà, L.}, \bibinfo{author}{Li, Z.}, \bibinfo{year}{2020}.
\newblock \bibinfo{title}{Multi-agent deep reinforcement learning for large-scale traffic signal control}.
\newblock \bibinfo{journal}{IEEE Transactions on Intelligent Transportation Systems} \bibinfo{volume}{21}, \bibinfo{pages}{1086--1095}.
\newblock \DOIprefix\doi{10.1109/TITS.2019.2901791}.
\bibitem[{Eom and Kim(2020)}]{Eom2020TrafficSignalControl}
\bibinfo{author}{Eom, M.}, \bibinfo{author}{Kim, B.I.}, \bibinfo{year}{2020}.
\newblock \bibinfo{title}{The traffic signal control problem for intersections: A review}.
\newblock \bibinfo{journal}{European Transport Research Review} \bibinfo{volume}{12}, \bibinfo{pages}{50}.
\newblock \URLprefix \url{https://doi.org/10.1186/s12544-020-00440-8}, \DOIprefix\doi{10.1186/s12544-020-00440-8}.
\bibitem[{Gu et~al.(2025)Gu, Wang, Jia, Zhang, Luo, Mao, Wang and Gee~Lim}]{10966978}
\bibinfo{author}{Gu, H.}, \bibinfo{author}{Wang, S.}, \bibinfo{author}{Jia, D.}, \bibinfo{author}{Zhang, Y.}, \bibinfo{author}{Luo, Y.}, \bibinfo{author}{Mao, G.}, \bibinfo{author}{Wang, J.}, \bibinfo{author}{Gee~Lim, E.}, \bibinfo{year}{2025}.
\newblock \bibinfo{title}{Communication strategy on macro-and-micro traffic state in cooperative deep reinforcement learning for regional traffic signal control}.
\newblock \bibinfo{journal}{IEEE Transactions on Intelligent Transportation Systems} \bibinfo{volume}{26}, \bibinfo{pages}{12183--12196}.
\newblock \DOIprefix\doi{10.1109/TITS.2025.3556931}.
\bibitem[{Gu et~al.(2024)Gu, Wang, Ma, Jia, Mao, Lim and Wong}]{10490249}
\bibinfo{author}{Gu, H.}, \bibinfo{author}{Wang, S.}, \bibinfo{author}{Ma, X.}, \bibinfo{author}{Jia, D.}, \bibinfo{author}{Mao, G.}, \bibinfo{author}{Lim, E.G.}, \bibinfo{author}{Wong, C.P.R.}, \bibinfo{year}{2024}.
\newblock \bibinfo{title}{Large-scale traffic signal control using constrained network partition and adaptive deep reinforcement learning}.
\newblock \bibinfo{journal}{IEEE Transactions on Intelligent Transportation Systems} \bibinfo{volume}{25}, \bibinfo{pages}{7619--7632}.
\newblock \DOIprefix\doi{10.1109/TITS.2024.3352446}.
\bibitem[{van Hasselt et~al.(2016)van Hasselt, Guez and Silver}]{vanHasselt2016Deep}
\bibinfo{author}{van Hasselt, H.}, \bibinfo{author}{Guez, A.}, \bibinfo{author}{Silver, D.}, \bibinfo{year}{2016}.
\newblock \bibinfo{title}{Deep reinforcement learning with double q-learning}, in: \bibinfo{booktitle}{Proceedings of the Thirtieth AAAI Conference on Artificial Intelligence (AAAI 2016)}, pp. \bibinfo{pages}{2094--2100}.
\newblock \DOIprefix\doi{10.5555/3016100.3016191}.
\bibitem[{Hu et~al.(2024)Hu, Li, Song, Xu, Xia, Sun, Zhou and Xia}]{HU2024128068}
\bibinfo{author}{Hu, K.}, \bibinfo{author}{Li, M.}, \bibinfo{author}{Song, Z.}, \bibinfo{author}{Xu, K.}, \bibinfo{author}{Xia, Q.}, \bibinfo{author}{Sun, N.}, \bibinfo{author}{Zhou, P.}, \bibinfo{author}{Xia, M.}, \bibinfo{year}{2024}.
\newblock \bibinfo{title}{A review of research on reinforcement learning algorithms for multi-agents}.
\newblock \bibinfo{journal}{Neurocomputing} \bibinfo{volume}{599}, \bibinfo{pages}{128068}.
\newblock \URLprefix \url{https://www.sciencedirect.com/science/article/pii/S0925231224008397}, \DOIprefix\doi{https://doi.org/10.1016/j.neucom.2024.128068}.
\bibitem[{{INRIX}()}]{inrix2024scorecard}
\bibinfo{author}{{INRIX}}, .
\newblock \bibinfo{title}{INRIX 2024 Global Traffic Scorecard}.
\newblock \bibinfo{type}{Tech. Rep.}. INRIX.
\newblock \URLprefix \url{https://inrix.com/scorecard/}.
\bibitem[{Li et~al.(2025)Li, Pan, Liu and Li}]{Li2025}
\bibinfo{author}{Li, M.}, \bibinfo{author}{Pan, X.}, \bibinfo{author}{Liu, C.}, \bibinfo{author}{Li, Z.}, \bibinfo{year}{2025}.
\newblock \bibinfo{title}{Federated deep reinforcement learning-based urban traffic signal optimal control}.
\newblock \bibinfo{journal}{Scientific Reports} \bibinfo{volume}{15}, \bibinfo{pages}{11724}.
\newblock \DOIprefix\doi{10.1038/s41598-025-91966-1}.
\bibitem[{Li et~al.(2024)Li, Zhang, Li and Sun}]{10637352}
\bibinfo{author}{Li, Y.}, \bibinfo{author}{Zhang, Y.}, \bibinfo{author}{Li, X.}, \bibinfo{author}{Sun, C.}, \bibinfo{year}{2024}.
\newblock \bibinfo{title}{Regional multi-agent cooperative reinforcement learning for city-level traffic grid signal control}.
\newblock \bibinfo{journal}{IEEE/CAA Journal of Automatica Sinica} \bibinfo{volume}{11}, \bibinfo{pages}{1987--1998}.
\newblock \DOIprefix\doi{10.1109/JAS.2024.124365}.
\bibitem[{Liang et~al.(2025)Liang, Du, Liu and et~al.}]{liang2025survey}
\bibinfo{author}{Liang, J.}, \bibinfo{author}{Du, X.}, \bibinfo{author}{Liu, Y.}, \bibinfo{author}{et~al.}, \bibinfo{year}{2025}.
\newblock \bibinfo{title}{A survey on multi-agent reinforcement learning for adaptive transportation solutions}.
\newblock \bibinfo{journal}{SN Computer Science} \bibinfo{volume}{6}, \bibinfo{pages}{955}.
\newblock \DOIprefix\doi{10.1007/s42979-025-04475-3}.
\bibitem[{Liang et~al.(2019)Liang, Du, Wang and Han}]{Liang2019}
\bibinfo{author}{Liang, X.}, \bibinfo{author}{Du, X.}, \bibinfo{author}{Wang, G.}, \bibinfo{author}{Han, Z.}, \bibinfo{year}{2019}.
\newblock \bibinfo{title}{A deep reinforcement learning network for traffic light cycle control}.
\newblock \bibinfo{journal}{IEEE Transactions on Vehicular Technology} \bibinfo{volume}{68}, \bibinfo{pages}{1243--1253}.
\newblock \DOIprefix\doi{10.1109/TVT.2018.2890726}.
\bibitem[{Liu and Ding(2022)}]{LIU2022390}
\bibinfo{author}{Liu, B.}, \bibinfo{author}{Ding, Z.}, \bibinfo{year}{2022}.
\newblock \bibinfo{title}{A distributed deep reinforcement learning method for traffic light control}.
\newblock \bibinfo{journal}{Neurocomputing} \bibinfo{volume}{490}, \bibinfo{pages}{390--399}.
\newblock \URLprefix \url{https://www.sciencedirect.com/science/article/pii/S092523122101818X}, \DOIprefix\doi{https://doi.org/10.1016/j.neucom.2021.11.106}.
\bibitem[{Liu et~al.(2025)Liu, Liu, Chen, Huang and Ding}]{LIU2025130834}
\bibinfo{author}{Liu, B.}, \bibinfo{author}{Liu, X.}, \bibinfo{author}{Chen, C.}, \bibinfo{author}{Huang, J.}, \bibinfo{author}{Ding, Z.}, \bibinfo{year}{2025}.
\newblock \bibinfo{title}{Decentralized neighboring information fusion for traffic network signal control}.
\newblock \bibinfo{journal}{Neurocomputing} \bibinfo{volume}{650}, \bibinfo{pages}{130834}.
\newblock \URLprefix \url{https://www.sciencedirect.com/science/article/pii/S0925231225015061}, \DOIprefix\doi{https://doi.org/10.1016/j.neucom.2025.130834}.
\bibitem[{Lopez et~al.(2018)Lopez, Behrisch, Bieker-Walz, Erdmann, Fl{\"o}tter{\"o}d, Hilbrich, L{\"u}cken, Rummel, Wagner and Wie{\ss}ner}]{sumo}
\bibinfo{author}{Lopez, P.A.}, \bibinfo{author}{Behrisch, M.}, \bibinfo{author}{Bieker-Walz, L.}, \bibinfo{author}{Erdmann, J.}, \bibinfo{author}{Fl{\"o}tter{\"o}d, Y.P.}, \bibinfo{author}{Hilbrich, R.}, \bibinfo{author}{L{\"u}cken, L.}, \bibinfo{author}{Rummel, J.}, \bibinfo{author}{Wagner, P.}, \bibinfo{author}{Wie{\ss}ner, E.}, \bibinfo{year}{2018}.
\newblock \bibinfo{title}{Microscopic traffic simulation using sumo}, in: \bibinfo{booktitle}{The 21st IEEE International Conference on Intelligent Transportation Systems}, \bibinfo{publisher}{IEEE}.
\newblock \URLprefix \url{https://elib.dlr.de/124092/}.
\bibitem[{Lu et~al.(2025)Lu, Li, Yu and Wang}]{LU2025}
\bibinfo{author}{Lu, Y.}, \bibinfo{author}{Li, C.}, \bibinfo{author}{Yu, H.}, \bibinfo{author}{Wang, H.}, \bibinfo{year}{2025}.
\newblock \bibinfo{title}{Soft actor-critic based regional traffic signal control in connected environment and its application in priority signal control}.
\newblock \bibinfo{journal}{Journal of Intelligent Transportation Systems} \URLprefix \url{https://www.sciencedirect.com/science/article/pii/S1547245025000325}, \DOIprefix\doi{https://doi.org/10.1080/15472450.2025.2532724}.
\bibitem[{Mnih et~al.(2015)Mnih, Kavukcuoglu, Silver, Rusu, Veness, Bellemare, Graves, Riedmiller, Fidjeland, Ostrovski, Petersen, Beattie, Sadik, Antonoglou, King, Kumaran, Wierstra, Legg and Hassabis}]{Mnih2015Human}
\bibinfo{author}{Mnih, V.}, \bibinfo{author}{Kavukcuoglu, K.}, \bibinfo{author}{Silver, D.}, \bibinfo{author}{Rusu, A.A.}, \bibinfo{author}{Veness, J.}, \bibinfo{author}{Bellemare, M.G.}, \bibinfo{author}{Graves, A.}, \bibinfo{author}{Riedmiller, M.}, \bibinfo{author}{Fidjeland, A.K.}, \bibinfo{author}{Ostrovski, G.}, \bibinfo{author}{Petersen, S.}, \bibinfo{author}{Beattie, C.}, \bibinfo{author}{Sadik, A.}, \bibinfo{author}{Antonoglou, I.}, \bibinfo{author}{King, H.}, \bibinfo{author}{Kumaran, D.}, \bibinfo{author}{Wierstra, D.}, \bibinfo{author}{Legg, S.}, \bibinfo{author}{Hassabis, D.}, \bibinfo{year}{2015}.
\newblock \bibinfo{title}{Human‐level control through deep reinforcement learning}.
\newblock \bibinfo{journal}{Nature} \bibinfo{volume}{518}, \bibinfo{pages}{529--533}.
\newblock \DOIprefix\doi{10.1038/nature14236}.
\bibitem[{Noaeen et~al.(2022)Noaeen, Naik, Goodman, Crebo, Abrar, Abad, Bazzan and Far}]{Noaeen2022}
\bibinfo{author}{Noaeen, M.}, \bibinfo{author}{Naik, A.}, \bibinfo{author}{Goodman, L.}, \bibinfo{author}{Crebo, J.}, \bibinfo{author}{Abrar, T.}, \bibinfo{author}{Abad, Z.}, \bibinfo{author}{Bazzan, A.}, \bibinfo{author}{Far, B.}, \bibinfo{year}{2022}.
\newblock \bibinfo{title}{Reinforcement learning in urban network traffic signal control: A systematic literature review}.
\newblock \bibinfo{journal}{Expert Systems with Applications} \bibinfo{volume}{199}, \bibinfo{pages}{116830}.
\newblock \DOIprefix\doi{10.1016/j.eswa.2022.116830}.
\bibitem[{Rasheed et~al.(2020)Rasheed, Yau, Noor, Wu and Low}]{Rasheed2020}
\bibinfo{author}{Rasheed, F.}, \bibinfo{author}{Yau, K.}, \bibinfo{author}{Noor, R.M.}, \bibinfo{author}{Wu, C.}, \bibinfo{author}{Low, Y.C.}, \bibinfo{year}{2020}.
\newblock \bibinfo{title}{Deep reinforcement learning for traffic signal control: A review}.
\newblock \bibinfo{journal}{IEEE Access} \bibinfo{volume}{8}, \bibinfo{pages}{208016--208044}.
\newblock \DOIprefix\doi{10.1109/ACCESS.2020.3034141}.
\bibitem[{Ren et~al.(2024)Ren, Dong, Zhao, Zhang, Kong and Yang}]{REN2024121111}
\bibinfo{author}{Ren, F.}, \bibinfo{author}{Dong, W.}, \bibinfo{author}{Zhao, X.}, \bibinfo{author}{Zhang, F.}, \bibinfo{author}{Kong, Y.}, \bibinfo{author}{Yang, Q.}, \bibinfo{year}{2024}.
\newblock \bibinfo{title}{Two-layer coordinated reinforcement learning for traffic signal control in traffic network}.
\newblock \bibinfo{journal}{Expert Systems with Applications} \bibinfo{volume}{235}, \bibinfo{pages}{121111}.
\newblock \URLprefix \url{https://www.sciencedirect.com/science/article/pii/S0957417423016135}, \DOIprefix\doi{https://doi.org/10.1016/j.eswa.2023.121111}.
\bibitem[{Saadi et~al.(2025)Saadi, Abghour, Chiba, Moussaid and Ali}]{saadi2025survey}
\bibinfo{author}{Saadi, A.}, \bibinfo{author}{Abghour, N.}, \bibinfo{author}{Chiba, Z.}, \bibinfo{author}{Moussaid, K.}, \bibinfo{author}{Ali, S.}, \bibinfo{year}{2025}.
\newblock \bibinfo{title}{A survey of reinforcement and deep reinforcement learning for coordination in intelligent traffic light control}.
\newblock \bibinfo{journal}{Journal of Big Data} \bibinfo{volume}{12}, \bibinfo{pages}{84}.
\newblock \URLprefix \url{https://journalofbigdata.springeropen.com/articles/10.1186/s40537-025-01104-x}, \DOIprefix\doi{10.1186/s40537-025-01104-x}.
\bibitem[{Saeedmanesh and Geroliminis(2016)}]{SAEEDMANESH2016250}
\bibinfo{author}{Saeedmanesh, M.}, \bibinfo{author}{Geroliminis, N.}, \bibinfo{year}{2016}.
\newblock \bibinfo{title}{Clustering of heterogeneous networks with directional flows based on “snake” similarities}.
\newblock \bibinfo{journal}{Transportation Research Part B: Methodological} \bibinfo{volume}{91}, \bibinfo{pages}{250--269}.
\newblock \URLprefix \url{https://www.sciencedirect.com/science/article/pii/S0191261515302605}, \DOIprefix\doi{https://doi.org/10.1016/j.trb.2016.05.008}.
\bibitem[{Song et~al.(2024)Song, Zhou and Ma}]{SONG2024104528}
\bibinfo{author}{Song, X.B.}, \bibinfo{author}{Zhou, B.}, \bibinfo{author}{Ma, D.}, \bibinfo{year}{2024}.
\newblock \bibinfo{title}{Cooperative traffic signal control through a counterfactual multi-agent deep actor critic approach}.
\newblock \bibinfo{journal}{Transportation Research Part C: Emerging Technologies} \bibinfo{volume}{160}, \bibinfo{pages}{104528}.
\newblock \URLprefix \url{https://www.sciencedirect.com/science/article/pii/S0968090X24000494}, \DOIprefix\doi{https://doi.org/10.1016/j.trc.2024.104528}.
\bibitem[{Spall(1992)}]{Spall1992}
\bibinfo{author}{Spall, J.C.}, \bibinfo{year}{1992}.
\newblock \bibinfo{title}{Multivariate stochastic approximation using a simultaneous perturbation gradient approximation}.
\newblock \bibinfo{journal}{IEEE Transactions on Automatic Control} \bibinfo{volume}{37}, \bibinfo{pages}{332--341}.
\newblock \DOIprefix\doi{10.1109/9.119632}.
\bibitem[{Sutton and Barto(2018)}]{SuttonBarto2018}
\bibinfo{author}{Sutton, R.S.}, \bibinfo{author}{Barto, A.G.}, \bibinfo{year}{2018}.
\newblock \bibinfo{title}{Reinforcement Learning: An Introduction}.
\newblock \bibinfo{edition}{2} ed., \bibinfo{publisher}{MIT Press}, \bibinfo{address}{Cambridge, MA}.
\bibitem[{Tan et~al.(2020)Tan, Bao, Deng, Jin, Dai and Wang}]{8676356}
\bibinfo{author}{Tan, T.}, \bibinfo{author}{Bao, F.}, \bibinfo{author}{Deng, Y.}, \bibinfo{author}{Jin, A.}, \bibinfo{author}{Dai, Q.}, \bibinfo{author}{Wang, J.}, \bibinfo{year}{2020}.
\newblock \bibinfo{title}{Cooperative deep reinforcement learning for large-scale traffic grid signal control}.
\newblock \bibinfo{journal}{IEEE Transactions on Cybernetics} \bibinfo{volume}{50}, \bibinfo{pages}{2687--2700}.
\newblock \DOIprefix\doi{10.1109/TCYB.2019.2904742}.
\bibitem[{Wang et~al.(2024)Wang, Zhu, Zhang, Tian and Zhang}]{WANG2024104582}
\bibinfo{author}{Wang, T.}, \bibinfo{author}{Zhu, Z.}, \bibinfo{author}{Zhang, J.}, \bibinfo{author}{Tian, J.}, \bibinfo{author}{Zhang, W.}, \bibinfo{year}{2024}.
\newblock \bibinfo{title}{A large-scale traffic signal control algorithm based on multi-layer graph deep reinforcement learning}.
\newblock \bibinfo{journal}{Transportation Research Part C: Emerging Technologies} \bibinfo{volume}{162}, \bibinfo{pages}{104582}.
\newblock \URLprefix \url{https://www.sciencedirect.com/science/article/pii/S0968090X24001037}, \DOIprefix\doi{https://doi.org/10.1016/j.trc.2024.104582}.
\bibitem[{Watkins and Dayan(1992)}]{Watkins1992}
\bibinfo{author}{Watkins, C.J.C.H.}, \bibinfo{author}{Dayan, P.}, \bibinfo{year}{1992}.
\newblock \bibinfo{title}{Q-learning}.
\newblock \bibinfo{journal}{Machine Learning} \bibinfo{volume}{8}, \bibinfo{pages}{279–292}.
\newblock \DOIprefix\doi{10.1007/BF00992698}.
\bibitem[{Wu et~al.(2020)Wu, Zhou, Liu, Yuan, Wang, Huang and Wu}]{9103316}
\bibinfo{author}{Wu, T.}, \bibinfo{author}{Zhou, P.}, \bibinfo{author}{Liu, K.}, \bibinfo{author}{Yuan, Y.}, \bibinfo{author}{Wang, X.}, \bibinfo{author}{Huang, H.}, \bibinfo{author}{Wu, D.O.}, \bibinfo{year}{2020}.
\newblock \bibinfo{title}{Multi-agent deep reinforcement learning for urban traffic light control in vehicular networks}.
\newblock \bibinfo{journal}{IEEE Transactions on Vehicular Technology} \bibinfo{volume}{69}, \bibinfo{pages}{8243--8256}.
\newblock \DOIprefix\doi{10.1109/TVT.2020.2997896}.
\bibitem[{Xing et~al.(2022)Xing, Li, Liu, Li and Zhang}]{su14169802}
\bibinfo{author}{Xing, Y.}, \bibinfo{author}{Li, W.}, \bibinfo{author}{Liu, W.}, \bibinfo{author}{Li, Y.}, \bibinfo{author}{Zhang, Z.}, \bibinfo{year}{2022}.
\newblock \bibinfo{title}{A dynamic regional partitioning method for active traffic control}.
\newblock \bibinfo{journal}{Sustainability} \bibinfo{volume}{14}.
\newblock \URLprefix \url{https://www.mdpi.com/2071-1050/14/16/9802}, \DOIprefix\doi{10.3390/su14169802}.
\bibitem[{Yang et~al.(2023)Yang, Zhang, Xu, Mao and Chen}]{YANG2023101431}
\bibinfo{author}{Yang, Q.}, \bibinfo{author}{Zhang, X.}, \bibinfo{author}{Xu, X.}, \bibinfo{author}{Mao, X.}, \bibinfo{author}{Chen, X.}, \bibinfo{year}{2023}.
\newblock \bibinfo{title}{Urban congestion pricing based on relative comfort and its impact on carbon emissions}.
\newblock \bibinfo{journal}{Urban Climate} \bibinfo{volume}{49}, \bibinfo{pages}{101431}.
\newblock \URLprefix \url{https://www.sciencedirect.com/science/article/pii/S2212095523000251}, \DOIprefix\doi{https://doi.org/10.1016/j.uclim.2023.101431}.
\bibitem[{Yau et~al.(2017)Yau, Qadir, Khoo, Ling and Komisarczuk}]{yau2017survey}
\bibinfo{author}{Yau, K.L.A.}, \bibinfo{author}{Qadir, J.}, \bibinfo{author}{Khoo, H.L.}, \bibinfo{author}{Ling, M.H.}, \bibinfo{author}{Komisarczuk, P.}, \bibinfo{year}{2017}.
\newblock \bibinfo{title}{A survey on reinforcement learning models and algorithms for traffic signal control}.
\newblock \bibinfo{journal}{ACM Computing Surveys} \bibinfo{volume}{50}.
\newblock \DOIprefix\doi{10.1145/3068287}.
\bibitem[{Yi et~al.(2022)Yi, Wu, Ren, Ran and Lou}]{9922459}
\bibinfo{author}{Yi, C.}, \bibinfo{author}{Wu, J.}, \bibinfo{author}{Ren, Y.}, \bibinfo{author}{Ran, Y.}, \bibinfo{author}{Lou, Y.}, \bibinfo{year}{2022}.
\newblock \bibinfo{title}{A spatial-temporal deep reinforcement learning model for large-scale centralized traffic signal control}, in: \bibinfo{booktitle}{2022 IEEE 25th International Conference on Intelligent Transportation Systems (ITSC)}, pp. \bibinfo{pages}{275--280}.
\newblock \DOIprefix\doi{10.1109/ITSC55140.2022.9922459}.
\bibitem[{Zhao et~al.(2024)}]{Zhao2024}
\bibinfo{author}{Zhao, H.}, et~al., \bibinfo{year}{2024}.
\newblock \bibinfo{title}{A survey on deep reinforcement learning approaches for traffic signal control}.
\newblock \bibinfo{journal}{Expert Systems with Applications} \bibinfo{volume}{234}, \bibinfo{pages}{119186}.
\newblock \DOIprefix\doi{10.1016/j.eswa.2024.2586}.

\end{thebibliography}





\end{document}